# NeuralPDE: Automating Physics-Informed Neural Networks (PINNs) with Error Approximations

Kirill Zubov*, Zoe McCarthy*, Yingbo Ma†,
Francesco Calisto‡, Valerio Pagliarino‡, Simone Azeglio‡§,
Luca Bottero‡, Emmanuel Luján*, Valentin Sulzer¶,
Ashutosh Bharambe‖, Nand Vinchhi, Kaushik Balakrishnan**,
Devesh Upadhyay**, Chris Rackauckas*†

July 19, 2021

**Abstract**

Physics-informed neural networks (PINNs) are an increasingly powerful way to solve partial differential equations, generate digital twins, and create neural surrogates of physical models. In this manuscript we detail the various methodologies of PINNs and showcase the various types of problems a PINN software can solve. We then detail the inner workings of NeuralPDE.jl and show how a formulation structured around numerical quadrature gives rise to new loss functions which allow for adaptivity towards bounded error tolerances. We describe the various ways one can use the tool, detailing mathematical techniques like using extended loss functions for parameter estimation and operator discovery, to help potential users adopt these PINN-based techniques into their workflow. We showcase how NeuralPDE uses a purely symbolic formulation so that all of the underlying training code is generated from an abstract formulation, and show how to make use of GPUs and solve systems of PDEs. Afterwards we give a detailed performance analysis which showcases the trade-off between training techniques on a large set of PDEs. From these comprehensive performance benchmarks we provide guidelines for users of PINN software, such as mixing ADAM optimizers with techniques like BFGS and mixing robust quadrature techniques with faster quasi-random samplers. We end by focusing on a complex multiphysics example, the

*Massachusetts Institute of Technology
†Julia Computing
‡University of Turin
§University of Ottawa
¶Carnegie Mellon University
‖Indian Institute of Technology, Roorkee
**Ford Motor Company

Doyle-Fuller-Newman (DFN) Model, and showcase how this PDE can be formulated and solved with NeuralPDE. Together this manuscript is meant to be a detailed and approachable technical report to help potential users of the technique quickly get a sense of the real-world performance trade-offs and use cases of the PINN techniques.

# 1 Introduction

Is it possible to solve a model at all possible parameters simultaneously, with an approximated error bound, in a way that does not exponentially grow in computational cost for high dimensions? The purpose of this manuscript is to demonstrate that this is not only possible but feasible using the latest techniques in scientific machine learning. Scientific machine learning is the burgeoning field combining techniques of machine learning into traditional scientific computing and mechanistic modeling. While scientific computing techniques generally can give clear measurable error bounds on predictions from interpretable models, machine learning is known to able to automatically adapt from data and, after being trained, can quickly give predictions of new phenomena. We will detail how numerical analysis techniques like adaptive quadrature can be employed to give physics-informed neural networks (PINNs) [46] the best of both of these properties. As such, the trained PINN will act as a neural surrogate with good analytical properties while giving fast predictions at new parameters. The PINN also includes ways to mix data into the physical model, building a digital twin capable of improving prediction accuracy. Together, this manuscript is a technical report detailing the mathematical techniques behind the NeuralPDE.jl software package, showcasing some new techniques for training PINNs, detailing different use cases for PINNs, and comparatively evaluating the performance of various PINN techniques.

We note that this is currently a living document with updates expected as NeuralPDE.jl includes more techniques. It is not intended to be a standard research document and is instead focused on being comprehensive to extensively document choices of hyper-parameters and which methods actually work well in practice. However, there are some novel aspects to make note of, such as the use of differentiable adaptive quadrature, with a comparative analysis against existing techniques.

# 2 Physics-Informed Neural Networks (PINNs)

## 2.1 The PINN Training Problem

The basis of our physical models will be partial differential equations (PDEs). Formally, a PDE can be defined as:

$$f(u, x; \lambda) = 0 \qquad (1)$$

for some nonlinear $f$ acting on $u(x)$ with $x \in \Omega$, where $\lambda$ are the parameters of the equation. For example, the classic heat equation can be written in this form as:

$$\frac{\partial u}{\partial t} - D\frac{\partial^2 u}{\partial x^2} = 0 \qquad (2)$$

where $D$ is the diffusion constant parameter. The goal is thus to find the function $u(t,x)$ which satisfies the equation over the chosen domain $\Omega$.

Physics-informed neural networks (PINNs) solve the PDE by using the Universal Approximation Theorem [46, 11, 23], which states that, informally, for any sufficiently regular function $u$, there is a large enough neural network $N$ with weights $w$ such that $\|N(x,w) - u(x)\| < \epsilon$ for all $x \in \Omega$. Given the known regularity results on PDEs [54], this suggests that an arbitrary PDE can be solved by replacing the unknown $u(x)$ with a neural network $N(x;w)$ into the constituent PDE and finding the weights $w$ such that $f(N;\lambda) \approx 0$ over all $x \in \Omega$. Formally, we can write this condition in one equation by summing up the difference at every point $x$,

$$C(w) = \int_\Omega \|f(N(x,w);\lambda)\| dx \qquad (3)$$

where we wish to find the weights of the neural network $w$ which minimize $C(w)$. This is the difference from the exact solution, where if $C(w) = 0$ then by definition the neural network is the solution to the differential equation.

To clarify the equation, we note that $f$ can be a vector of possible equations. For example, take partial differential-algebraic equation system (PDAE) [35]:

$$\frac{\partial u_1}{\partial t} = \frac{\partial^2 u_1}{\partial x^2} + u_3 \sin(\pi x) \qquad (4)$$

$$\frac{\partial u_2}{\partial t} = \frac{\partial^2 u_2}{\partial x^2} + u_3 \cos(\pi x) \qquad (5)$$

$$0 = u_1 \sin(\pi x) + u_2 \cos(\pi x) - e^{-t} \qquad (6)$$

with initial conditions:

$$u_1(0,x) = \sin(\pi x) \qquad (7)$$
$$u_2(0,x) = \cos(\pi x) \qquad (8)$$
$$\frac{\partial u_1(0,x)}{\partial t} = -\sin(\pi x) \qquad (9)$$
$$\frac{\partial u_2(0,x)}{\partial t} = -\cos(\pi x) \qquad (10)$$
$$\qquad (11)$$

For this set of 7 equations, define the solution via neural networks $N_i = u_i$ (or via a single neural network with vector output $N(x) = [u_i]$), where $f$ outputs a vector of 7 values which includes the algebraic restrictions and the boundary conditions. However, as the boundary conditions only need to be satisfied on (some subset of) $\partial\Omega$, it can be helpful to instead split the boundary terms into their own equation. If we separate $f$ into its component functions, we thus receive:

$$C(w) = \sum_i \int_{\Omega \setminus \partial\Omega} \|f_i(N(x,w);\lambda)\| dx + \sum_i \int_{\partial\Omega} \|b_i(N(x,w);\lambda)\| dx \qquad (12)$$

where $f_i$ is each equation in the system of PDEs (including the algebraic equations) and $b_i$ are the boundary conditions. While technically equivalent to Equation 3, we will refer to the terms in the form of Equation 12 as it clarifies the implementation.

## 2.2 Using PINNs to Generate Neural Surrogates

One of the advantages of using a machine learning architecture like neural networks is that they overcome the curse of dimensionality [2, 5]. In a concrete sense, the curse of dimensionality is the property of many algorithms to increase their computational cost exponentially when the dimensionality of the problem is increased. For example, when a finite difference discretization is used, $N$ points in every dimension $d$ requires $N^d$ points. To intuitively understand the effect of this exponential growth, a 3-dimensional partial differential equation solved using 1000 points along each axis with a finite difference method would take $1000^3 \times 64\text{bits} \approx 8\text{GB}$ of RAM to hold the solution, which is possible in a high end laptop, while for a 4-dimensional partial differential equation it would require $1000^4 \times 64\text{bits} \approx 8\text{TB}$ of RAM which is only feasible on specialized nodes of a high-performance compute cluster, while the 5-dimensional version is completely infeasible with modern technology. On the contrary, using the PINN technique we simply define a $N : R^5 \to R$ neural network parameterized by weights $w$. A mixture of theory and practice suggests that the weights required to approximate such a higher dimensional function grows only polynomially with this input dimension [19, 24], and one of the most common examples in machine learning is the MNIST dataset which requires training neural networks on 784-dimensional inputs ($28 \times 28$ pixel representations of images) [12].

While this advantage may seem moot given the physical world is 3-dimensional, this ability to handle high dimensional can be utilized in many real-world scenarios. For example, one case where high dimensional PDEs naturally arise is in the evolution of probability distributions, where every object being considered gives rise to another dimension in the partial differential equation [44]. For example, a nonlinear Black-Scholes equation predicting the valuation of a portfolio of $d$ options requires solving a $d$-dimensional partial differential equation [22]. Similarly, solutions to optimal controls of stochastic dynamical systems are given by the Hamilton-Jacobi-Bellman equation which is a $d$-dimensional PDE, where $d$ is the number of parameters used to describe the controller. Thus while physical PDEs are 3 dimensions or less, mathematical descriptions of various phenomena can still require solving arbitrarily high-dimensional PDEs where neural network techniques have found use.

One case where the high dimensional property of PINNs can be useful is in the "neural surrogatization" of PDE solutions. A surrogate model is an approximate model which allows for quickly receiving accurate enough results to perform some real-world engineering task. For example, in nuclear physics one can be develop high-fidelity models which require months to evaluate, while a surrogate of said model can be trained on previously generated model results to guess the model's prediction at new parameters in orders of magnitude less time. However, one of the disadvantages of surrogate models is that it can be difficult to estimate the error of the surrogate model. Using a trick, any error estimation technique for a PINN naturally extends to the surrogate.

The trick to using PINNs for neural surrogatization is to represent "parameters" as independent variables. For example, in the diffusion equation (Equation 60) one could instead think of the diffusion constant $D$ as an alternative independent variable, changing the equation to a 3-dimensional partial differential equation where each slice along the $D$ axis represents the solution to the 2-dimensional partial differential equation. Thus training the neural network to be the PDE solution in this form gives a network that can predict the solution at new parameters, reducing the total computational cost if one wishes to investigate a high-dimensional design space.

## 2.3 Discretizing the Loss Function and Obtaining Error Bounds

Evaluating the loss function of a PINN is equivalent to approximating the loss function of Equation 3, which computationally is most efficiently done by splitting the loss into its component pieces of Equation 12 and integrating over only the relevant domains for each portion. For example, the boundary conditions can be thought of as 0 at all values $x \in \Omega \backslash \partial\Omega$, but given this one should computationally only evaluate them on $x \in \partial\Omega$, making the integrals for the boundary terms lower dimensional than the PDE terms. Likewise, the difference in dimension is usually only one, and so while the performance of different integration strategies can greatly differ due to dimension, in practice the same technique can be used on the boundary and PDE terms, though with less points taken on the boundary to compensate for the change in the dimensionality. We note that this will naturally give more weight to the PDE terms, potentially causing the optimization to ignore the boundary terms, leading to the reweighing strategies discussed in Section 2.10.

Thus given one of the integral terms, we wish to find a set of points $x_i$ on which to evaluate the loss function. The simplest approach for training a PINN is to pick a set of points $x_i$ on a grid and approximating the integral via the Trapezoidal rule:

$$\int_{\Omega \backslash \partial\Omega} \|f_i(N(x,w);\lambda)\|dx \approx \sum_i \Delta x \|f(x_i)\| \tag{13}$$

We note that this is a modification of the normal Trapezoidal rule as the standard form includes $\frac{1}{2}f(x_i)$ on the boundary terms, which are omitted from this form. This is exposed as the `GridTraining` strategy within NeuralPDE.jl. The downside to this approach however, is that the number of grid points scales as $N^d$ for $N$ points along each dimension, and thus this formulation suffers from the curse of dimensionality for the same reasons as the finite difference methods. Worse, the neural network is never evaluated between the grid points, which might overfit the training data. This leads the neural network to evaluate a low loss while being unruly between the training points. Thus in many cases we turn to alternative losses.

One of the simplest ways to overcome the "unruly between training points" issue is to randomly stagger the training points. One way this can be done is simply by taking $x_i$ at a fixed number of random points and using the Monte Carlo integration method:

$$\int_{\Omega\setminus\partial\Omega} \|f_i(N(x,w);\lambda)\| dx \approx \alpha \sum_i \|f(x_i)\| \qquad (14)$$

For the correct constant $\alpha$, this converges to the true integral at a rate of $\mathcal{O}(\sqrt{N})$. However, given we wish to minimize this loss to zero, we can ignore the calculation of this constant and use the summation itself as the loss value, giving rise to the `StochasticTraining` strategy in NeuralPDE.jl. However, this low rate of convergence can sometimes hamper the method, requiring a large $N$ to actively remedy poor performance of the neural network in parts of the domain. In order to improve the spacing between randomly chosen points, one can use a low discrepancy sequence, also known as a quasi-Monte Carlo sample, to arrive at a convergence rate of $\mathcal{O}(N)$ [41, 28, 37]. Popular low discrepency sampling techniques, such as Sobol sequences [50, 25] and Latin Hypercubes [53, 30], are implemented in `QuasiMonteCarlo.jl` implemented as the `QuasiRandomTraining` strategy in NeuralPDE.jl.

Thus far, all of the aforementioned techniques used a fixed number of points $N$. However, adaptive quadrature techniques can be used which choose $N$ based on a given error estimate. In one dimension, the common techniques for this are Gauss-Kronrod quadrature [27] and Clenshaw-Curtis quadrature [58], where the latter uses a transformation of the Fourier basis and thus converges better for smooth functions while the former has more optimality on less regular functions [55]. In higher dimensions, these methods correspond to the cubature techniques [52, 51], where h-cubature methods are the higher dimensional analogue of the Gauss-Kronrod quadratures and p-cubature methods are the analogue of the Clenshaw-Curtis techniques [38]. These techniques are well-established to be more efficient than (quasi-)Monte Carlo methods for dimensions less than around 6 to 8, but fall off at higher dimensions due to curse of dimensionality. For higher dimensional adaptive quadratures, adaptive Monte Carlo techniques like VEGAS [29] and MISER [49] adaptively improve the weighing schemes to account for the most variation in the loss with the least points using successive evaluations of the integrand. While these receive the lower convergence rate of $\mathcal{O}(\sqrt{N})$ (and $\mathcal{O}(N)$ for adaptive quasi Monte Carlo), they can be advantagous to use over the non-adaptive versions due to both convenience and by greatly reducing the constant factor. These quadrature techniques have many pre-existing implementations in Julia in the Cuba.jl [20], Cubature.jl [38], and HCubature.jl libraries, which are all wrapped into Quadrature.jl [45]. Using these quadratures to evaluate the loss integrand is implemented as the `QuadratureTraining` strategy, which takes as an argument the choice of quadrature method.

One natural advantage of the adaptive quadrature techniques is that, since they all include some form of error approximation calculated as part of the adaptivity scheme [40, 1, 39], using the `QuadratureTraining` gives a measure of the error of the integrals, and thus and error estimate of Equation 3. This thus solves the earlier mentioned problem or neural surrogatization, where in this form the error bounds of the neural network's prediction at unknown (or all) parameters can be computed by summing the error bounds from the individual quadrature estimates. This gives a total expected error on the PDE definition, which can be shown to be proportional to the error on the dependent variables $u$.

### 2.4 Differentiating the Quadrature Loss Function

For most of the quadrature strategies, differentiation is straightforward as the discretized loss is given as simply $C(w) = \sum_i \alpha_i C(w, x_i)$, i.e. the summation of the cost at given points in the domain summed with respect to some constants $\alpha_i$. Due to the chain rule,

$$\frac{\partial C}{\partial w} = \sum_i \alpha_i \frac{\partial C(w, x_i)}{\partial w} \qquad (15)$$

which will naturally be computed by any competent automatic differentiation library. However, special care has to be taken with the adaptive quadrature techniques. If we use a black box adaptive quadrature method, like Cuba from C, then automatic differentiation libraries like Zygote will not be able to naively differentiate the function [43]. There are two possible ways to differentiate such an adaptive quadrature. Given sufficient regularity assumptions, the differentiation of the quadrature is done by exchanging the limits, that is:

$$\frac{\partial}{\partial w} \int_{\Omega \setminus \partial \Omega} g(x, w) dx = \int_{\Omega \setminus \partial \Omega} \frac{\partial g(x, w)}{\partial w} dx \qquad (16)$$

where $g(x, w) = \|f_i(N(x, w); \lambda)\|$. However, given that the points for evaluating the integral are dependent on $g$ itself (or stochastic), one can imagine two separate ways to compute said integral. If one differentiates the summation used to calculate the original cost, then one is enforcing that the derivative calculation discretizes Equation 16 at precisely the same points as the original cost function. The alternative is to treat the derivative integral as its own calculation and calculate the derivative at alternative points while still approximating Equation 16. The former approach is known as Discretize-then-Optimize (DtO) while the latter is Optimize-then-Discretize (OtD), also referred to as the discrete sensitivity and the continuous sensitivity. In other applications like differential equations and control there is a well-established trade-off between the two, with DtO generally being more stable with OtD being more computationally efficient, but with counter examples in each case [44]. That said, for the purposes at hand, implementing DtO would require the ability to have the quadrature library return the points $x_i$ at which the integration was calculated at, and thus the `QuadratureTraining` implements the derivative using the OtD derivative by defining forward and reverse mode rules which evaluate the integral of the derivative of the integrand. The other strategies use the natural DtO interpretation.

## 2.5 Imposing Constraints in PINN Solutions

One advantage of the PINN formulation is the natural ability to enforce constraints on the solution. This kind of equation is also known as a partial differential-algebraic equation (PDAE) and was shown as an example Equation 4. There are two ways to enforce such constraints: either through the loss function or structurally. In the loss function form, the PDAE terms simply arise as a term $f_i$ to be integrated and added to the loss, and thus when the loss is minimized, if this term is near zero, then the equality is approximately enforced. We note that boundary conditions are treated equivalently to differential-algebraic equations through this form. One can also use constrained optimization techniques to further enforce such constraint terms.

The other way to impose constraints on the PINN solutions is structurally by choosing an architecture which requires the condition to be satisfied [26]. For example, if one wished to solve an ODE via the PDE form, the only boundary condition would be the initial condition of the ODE $u(0) = u_0$. In this case, instead of choosing an arbitrary neural network $N$ as the function approximator, we can choose the function approximator:

$$\psi(t) = tN(t) + u_0 \tag{17}$$

which by design has $\psi(0) = u_0$ and thus no extra loss terms are required for that boundary condition. We note that enforcing such constraints in this form can be more efficient and stabilize the training, and thus NeuralPDE.jl's symbolic form allows for arbitrary Julia functions to serve as problem-informed non-universal approximators.

## 2.6 Estimating Differential Equation Parameters and Operator Discovery with PINNs

In many cases the physical parameters $\lambda$ of the equation can be unknown and need to be found through fitting to data. To do this, the optimization process can jointly optimize the neural network weights $w$ with the physical parameters $\lambda$ [31]. In the case a function needs to be learned, such as a spatially dependent diffusion constant a la:

$$\frac{\partial u}{\partial t} - D(x)\frac{\partial^2 u}{\partial x^2} = 0 \tag{18}$$

this spatial $D(x)$ can be learned simultaneously to the solving process by representing $D(x)$ as a neural network $M(x, v)$ with weights $v$, and learning the weights $v$ simultaneously to the solution weights $w$. Similarly, unknown operators of the equation can be recovered by representing the operators as nonlinear operators of the solution, i.e. $D(u, x)$, and similarly representing these operators as neural networks and appending the weights. This gives the ability to automatically discover and represent unknown operators and weights.

## 2.7 Incorporating Non-Local Operators into PINNs: Integro-Differential Equations and Fractional Differential Equations

In the previous sections, we discussed treatment of differential operators. In this section we show how PINNs can also be applied to equations with other types of operators, specifically non-local operators like the integral operator and fractional derivative operator. These operators are discretized using other methods such as quadrature algorithms or finite differences.

Let us consider an Integro-Differential Equations (IDEs) such as,

$$\frac{du(x)}{dx} + g(x, u(x)) = \int_{x_0}^{x} f(t, u(t)) dt \tag{19}$$

As we can see we can still use automatic differentiation for the derivative, but in order to calculate the integral we use numerical methods to approximate the integral function. Because the neural network defines the value of $u(x)$ over all space, one can simply use a quadrature technique, such as those in `Quadrature.jl` , to evaluate the current value of the integral and thus utilize these terms within the PDE equations. Similarly to any standard term, this then contributes to the loss function which causes the optimization process to minimize the loss and thus solve the integral terms.

Another non-local operator of interest are fractional derivatives. We can also extend PINNs to Fractional Partial Differential Equation (FPDEs) such as,

$$\mathcal{L}u(x,t) = f(x,t) \tag{20}$$
$$u(x,0) = g(x) \tag{21}$$
$$u(x,t) = 0 \tag{22}$$

When $\mathcal{L} = \frac{\partial^\gamma}{\partial t^\gamma} + c(-\Delta)^{\frac{\alpha}{2}} + \nu\nabla$ we can express the fractional derivative as an integral,

$$D^\alpha f(x) = \frac{1}{\Gamma(1-\alpha)} \frac{d}{dx} \int_0^x \frac{f(x)}{(x-t)^\alpha} \tag{23}$$

and thus as discussed, we can use numerical methods to calculate the fractional derivative after expressing it in the form of the integral. We can also define fractional derivatives using finite difference schemes as follows [42],

$$\frac{\partial u(x,t)^\gamma}{\partial t^\gamma} = \frac{1}{\Gamma(2-\gamma)(\Delta t)^\gamma} \left\{ -c_{\lceil \lambda t \rceil -1} u(x,0) + c_0 u(x,t) \right.$$
$$\left. + \sum_{k=1}^{\lceil \lambda t \rceil -1} (c_{\lceil \lambda t \rceil -k} - c_{\lceil \lambda t \rceil -k-1}) u(x, k\Delta t) \right\}$$

Here $c_l = (l+1)^{1-\gamma} - l^{1-\gamma}$. And $\Delta t = \frac{t}{\lceil \lambda t \rceil} \approx \frac{t}{\lambda t}$

We can again employ finite difference scheme for the fractional Laplacian. We start by approximating the directional fractional derivative using the shifted vector Grunwald-Letnikov method.

$$D_\theta^\alpha u(x,t) = \frac{1}{\Delta x^\alpha} \sum_{k=1}^{\lceil \lambda d(x,\theta,\Omega) \rceil} (-1)^k \binom{\alpha}{k} u(x - (k-1)\Delta x \theta, t) + O(\Delta x) \quad (24)$$

Here $\theta$ is the differentiation direction, $\Omega$ is the bounded domain of x, and $d(x,\theta,\Omega)$ is the distance of x from the boundary of the domain $\Omega$ in $-\theta$ direction and the spatial step $\Delta x \approx \frac{1}{\lambda}$ And then,

$$(\Delta)^{\frac{\alpha}{2}} u(x,t) = C_{\alpha,D} \int_{||\theta||^2} D_\theta^\alpha u(x,t) d\theta \quad (25)$$

and here, $C_{\alpha,D} = \frac{\Gamma(\frac{1-\alpha}{2})\Gamma(\frac{D+\alpha}{2})}{2\pi^{\frac{D+1}{2}}}$

This technique is not currently in NeuralPDE.jl but is scheduled to be implemented soon.

## 2.8 Incorporating Noise: Solving Stochastic PDEs with PINNs

While considering PINNs for stochastic differential equations (SDEs) we have to take into account another source of uncertainty: *parametric uncertainty*, which is introduced by the random parameter in the equation, represented as a stochastic process. For this kind of uncertainty quantification we need a robust data-driven technique for building the orthogonal basis for the random space, without any prior assumption on the distribution of the data. As shown by Zhang et al. [60], an approach which combines *arbitrary polynomial chaos* (aPC) with PINNs, named NN-aPC, where the NN is used to learn individual modes of the aPC expansion, is considered. The parametric uncertainty is taken together with another source of uncertainty which is due to the NN itself and is usually referred to as *approximation uncertainty*. This last additional source of uncertainty has been traditionally estimated by employing Bayesian Neural Networks (BNNs) which are time-consuming due to the fact that inference is not generally tractable. Dropout, one of the most used regularization techniques, has been recently adopted to measure uncertainty[17]. The main advantages of introducing dropout layers are their reduced computational cost and the fact that they can be easily combined with physical laws. Dropout can thus be used to estimate the uncertainty in approximating each aPC mode [60].

Let us now consider a stochastic partial differential equation, which can be written as follows:

$$\mathcal{N}[u(x,\lambda), k(x)] = 0 \tag{26}$$

where $\mathcal{N}$ is a nonlinear operator and the random parameter $k(x)$ is the parametric uncertainty source. The PINN approach generalization for solving SDEs can be achieved by applying the previously mentioned NN-aPC method which consists of three steps:

- dimensionality reduction by employing principal component analysis (PCA) to extract a set of uncorrelated random variables, which are the coordinates of the quantity of interest in the stochastic space;

- building the aPC basis, which is purely data-driven and the orthogonality holds by employing a discrete measure approximation of the underlying probability distribution of multiple measurements on the system;

- constructing the NN-aPC as a surrogate for aPC modes and train this network for each mode

For a more detailed description of the first step, let us consider the $N_k \times N_k$ covariance matrix for the measurements on $k$:

$$K_{i,j} = Cov[k_j, k_i] \tag{27}$$

From which, by applying PCA we get:

$$K = \Phi^T \Lambda \Phi \tag{28}$$

where, as usual, $\Phi$ is an orthonormal matrix and $\Lambda$ is a diagonal matrix. Let us measure our random parameter at time $s$, this yields to a vector $\mathbf{k_s}$. We can write down an uncorrelated random vector:

$$\boldsymbol{\xi_s} = \Phi^T \sqrt{\Lambda}^{-1} \mathbf{k_s} \tag{29}$$

hence, $\mathbf{k_s}$ can be written by means of a reduced M-dimensional ($M < N_k$) expansion:

$$\mathbf{k_s} \approx \mathbb{E}[\mathbf{k}] + \sqrt{\Lambda^M} \Phi^M \boldsymbol{\xi_s^M} \tag{30}$$

where $\mathbb{E}[\mathbf{k}]$ is the expectation of each measurement.

The second step is indeed centered around aPC. By recovering the uncorrelated random vector, we can assume that we have a set of random vectors $S := \{\boldsymbol{\xi_s}\}_{s=1}^N$ with unknown probability measure $\rho(\boldsymbol{\xi})$. In the regime where we have taken a large number of snapshots of our system, i.e. measurements at fixed time for several $t$s, it is feasible to approximate $\rho(\boldsymbol{\xi})$ by the discrete measure $\nu_S(\boldsymbol{\xi})$ which can be represented as a combination of Dirac measures:

$$\rho(\boldsymbol{\xi}) \approx \nu_S(\boldsymbol{\xi}) = \frac{1}{N} \sum_{\boldsymbol{\xi_s} \in S} \delta_{\boldsymbol{\xi_s}}(\boldsymbol{\xi}) \tag{31}$$

After that, a set of orthonormal polynomial basis functions $\{\Psi_\alpha(\boldsymbol{\xi})\}$ can be constructed by using the following recursive algorithm:

$$\Psi_\alpha(\boldsymbol{\xi}) = w_\alpha^\alpha \prod_{i=1}^{M} \boldsymbol{\xi}_{\mathbf{i}}^{\alpha_i} - \sum_{\beta \prec \alpha} w_\beta^\alpha \Psi_\beta(\boldsymbol{\xi}) \tag{32}$$

where $w_\beta^\alpha$ coefficients are determined by imposing the orthonormality condition with respect to the discrete measure $\nu_S$. By exploiting the polynomial basis $\{\Psi_\alpha(\boldsymbol{\xi})\}$ we can express any function of choice $g(x;\boldsymbol{\xi})$ as an aPC expansion:

$$g(x;\boldsymbol{\xi}) = \frac{1}{N} \sum_{\alpha=0}^{\frac{(r+M)!}{r!M!}-1} \sum_{s=1}^{N} \Psi_\alpha(\boldsymbol{\xi_s}) g(x;\boldsymbol{\xi_s}) \Psi_\alpha(\boldsymbol{\xi}) \tag{33}$$

where $r$ represents the highest polynomial order in $\Psi_\alpha(\boldsymbol{\xi})$.

Regarding the last step, we consider the inverse problem and we set up two NNs:

- the network $\hat{u}_\alpha$ which takes $x$ as the input and outputs a $\frac{(r+M)!}{r!M!} \times 1$ vector of the aPC modes of u evaluated in $x$;

- the network $\hat{k}_i$ that takes the coordinate $x$ as the inputs and outputs a $(M+1) \times 1$ vector of the modes of $k$.

We can approximate $k$ and $u$ by employing (30) and (33). The loss function is the sum of the mean squared errors (MSEs).

This technique is not currently in NeuralPDE.jl but is scheduled to be implemented soon.

## 2.9 Generating Digital Twins with Extended Loss Functions

In order to fit such data, one needs to extend the loss function to include terms relating to the data. This can be done by simply appending standard loss functions to the PINN loss. For example, the common L2 loss against some discrete dataset $d_i$ at points $x_i$ can be added to Equation 3 as:

$$C(w) = \int_\Omega \|f(N(x,w);\lambda)\| dx + \sum_i \|N(x_i,w) - d_i\| \tag{34}$$

In this form, one can think of the PINN loss as a regularization of the standard supervised training process, where said regularization pushes the solution towards an assumed physical relation, and thus the name physics-informed neural network. This can allow for the PINN solution to not necessarily follow the PDE solution when data suggests otherwise, giving rise to a digital twin which merges physical principles with data. Additionally, this loss function extension allows for parameter estimation as in Section 2.6 to make use of data. NeuralPDE.jl thus allow for the user to define and arbitrary function of the neural network to extend the PINN loss to include such data losses to allow for these applications.

## 2.10 Adaptive Reweighing of Loss Function Terms

The solution of a PDE is the simultaneous satisfaction of several equations: the initial conditions, boundary conditions, and internal domain PDE equations, and possibly other constraints. When we solve a PDE by training a PINN, each of these equations are transformed into a loss function $f_i$ or $b_i$, and then weighted by some factor, and added together to form a final loss function

$$C(w) = \sum_i \alpha_i C_{r_i}(w, x_i) + \sum_j \alpha_j C_{b_j}(w, x_j)$$

that is minimized to train a neural network for PDE loss terms $C_{r_i}$ and boundary condition (BC) loss terms $C_{b_j}$. When these equations are transformed into loss functions for this process, the scale of each loss function may be considerably different, as the PDE constraint equations can be arbitrary and could contain derivatives of many scales, large or small constants, and many other factors. For example, if we take a Fourier decomposition of a function $f(x)$ into a sum or integral of $\sin(\omega_i x)$ and $\cos(\omega_i x)$ factors: let $f(x) = \sum_i \beta \sin(\omega_i x)$ for some constants $\beta$ and $\omega_i$ for simplicity, but an extension of this analysis applies for more complicated Fourier decompositions. Each derivative multiplies the scale of the resultant component function by $\omega_i$:

$$\frac{\partial f(x)}{\partial x} = \sum_i \frac{\partial \beta \sin(\omega_i x)}{\partial x} = \sum_i \beta \omega_i \cos(\omega_i x) \tag{35}$$

Since every function admits a Fourier decomposition of some kind, derivatives will have this frequency scale amplification. The internal domain PDE equations tend to have more derivatives as regularizers or constraints than the boundary condition equations, so they are typically subject to more of this scale amplification. In the PINN case, we're learning neural net functions $f(x)$. Common initializations for neural networks tend to be fairly linear initially and thus have low derivative frequency scale amplification in the component loss functions. If the desired/true solution $u$ has high frequency in regions of the domain, then as a successful optimization progresses, the scale differences of the component loss functions can get more pronounced and change relative to one another. These derivative frequency scale differences are just one of several ways that the component loss functions have different scales, and these scale differences lead to many optimization difficulties while training the PINN.

There has been a surge of analysis in the literature on this difficulty in optimization recently. A common thread throughout most of these analyses is that they consider solutions that are focused on selecting the appropriate weights $\alpha_i$ for the combination of the loss terms. These $\alpha_i$ can then be carefully selected constants, or they can be adaptive by tracking some statistics about the training process.

One lens to understand this optimization issue is through the analysis of [59], where the authors analyze the training of PINNs as the solution of a stiff ODE, where the stiffness results from the scaling differences between the loss terms, and in particular the scale differences between the gradients of the PDE loss terms and the boundary condition loss terms. They suggest an adaptive $\alpha_i$ that is inspired by the ADAM optimizer. Their algorithm calculates $\alpha_i$ by an exponential moving average of the ratio between the maximum of the absolute values of the gradients of the PDE loss terms divided by the mean of the absolute values of the gradients of the boundary condition loss terms:

$$\hat{\alpha}_i = \frac{\max_\theta\{|\nabla_\theta C_r(\theta_n)|\}}{mean_\theta\{|\nabla_\theta C_b(\theta_n)|\}} \qquad (36)$$

and $\alpha_i = (1-\gamma)\lambda_i + \gamma\hat{\alpha}_i$ is the adaptive weight used in the loss function sum for each training iteration. This reweighting is performed to make the gradients of the boundary condition loss terms in approximately the same scale as the gradients of the pde loss terms. This adaptation can be performed as often as every optimization step or a set number of steps, such as every 10 optimization steps. In NeuralPDE, there is an adaptation of this method to apply to PDEs with multiple PDE loss terms and BC loss terms, since the original formulation only assumes one PDE loss term. We remove the max numerator term in the weighting, since that presupposes the selection of one special PDE loss term, and have the weights for each loss term be set before the exponential moving average by only the division of the mean of its loss gradient. We will refer to this method as the Loss Gradients adaptive reweighting method, because the reweighting is performed using the gradients of the losses.

If adaptivity in the weights is not included, this analysis can help to assign constants $\alpha_j$ for the BC loss terms that are much larger than the $\alpha_i$. There are several lines of research that analyse the PDE to be solved in order to derive optimal constant weights for the loss terms in the PINN loss function for that pde. In NeuralPDE these analytical weight selection methods are not implemented but the user can provide the $\alpha$ constants to the optimization manually.

Another way to select and adapt the weight terms for the loss terms is to reformulate the PINN training process as a minimax optimization [36]:

$$\min_w \max_{\alpha_{r_i},\alpha_{b_j}} C(w, \alpha_{r_i}, \alpha_{b_j})$$

the authors interpret this as a form of soft attention, which is a neural network architectural feature that has seen widespread success recently [57]. The inner maximization is performed via gradient ascent, and in this case since each $\alpha_i$ enters into the cost function only multiplied by its respective loss term $C_i(w)$, the gradient with respect to each $\alpha_i$ is simply the loss term it corresponds to: $\alpha_{i_n} = \alpha_{i_{n-1}} + \nu C_i(w)$. This can also be performed using a more advanced optimization procedure such as ADAM. Since these quantities are computed every iteration of minimization, this maximization can be performed at the same time as the outer minimization or less often. This allows the each $\alpha_i$ terms to automatically become larger as the loss term is not being satisfied during optimization. There is still an issue of scale, as an $\alpha_i$ term whose loss term has a high scale will tend to have the $\alpha_i$ grow faster with this gradient ascent. This can be somewhat mitigated by having a separate learning rate for the gradient ascent for PDE loss $\alpha_{r_i}$ and for the BC loss $\alpha_{b_j}$, where the learning rate for the BC loss $\alpha_{b_j}$ will be much higher than the learning rate for the $\alpha_{r_i}$. In NeuralPDE and the rest of this paper, we will refer to this method as the MiniMax adaptive reweighting method.

## 3 Automating the Symbolic Generation of PINN Losses

### 3.1 NeuralPDE Overview

While other PINN libraries require the user to write a low level formulation of the partial differential equation, NeuralPDE abstracts the implementation away from the structure by using a symbolic interface (defined via Symbolics.jl [18]) for defining the PDEs. The formal specification of the PDE is done via the `PDESystem` structure of ModelingToolkit.jl [33]. This is a SciML organization wide interface, and thus multiple packages define `discretize` and `symbolic_discretize` dispatches which transform the PDE specification into various numerically solvable forms. For exmaple, DiffEqOperators.jl lowers the equations via finite difference and finite volume forms into ODE systems using the method of lines in order for it to be solved by DifferentialEquations.jl [45]. In this sense, the interface for NeuralPDE is to offer a dispatch of `discretize` which takes in the `PDESystem` and returns a GalacticOptim.jl `OptimizationProblem` which can then be solved by numerical optimizers.

### 3.2 Simple Poisson Example

The symbolic form is as follows. We note that specifics of the syntax are subject to change and one should consult the documentation for the latest version. To specify the Poisson equation:

$$\frac{\partial^2 u}{\partial x^2} + \frac{\partial^2 u}{\partial y^2} = -\sin(\pi x)\sin(\pi y) \tag{37}$$

with boundary conditions:

$$u(0, y) = 0 \tag{38}$$
$$u(1, y) = -\sin(\pi)\sin(\pi y) \tag{39}$$
$$u(x, 0) = 0 \tag{40}$$
$$u(x, 1) = -\sin(\pi x)\sin(\pi) \tag{41}$$

on the domain $x \in [0, 1]$ and $y \in [0, 1]$, one would write:

```
using ModelingToolkit

@parameters x y
```

```julia
@variables u(..)
Dxx = Differential(x)^2
Dyy = Differential(y)^2

# 2D PDE
eq  = Dxx(u(x,y)) + Dyy(u(x,y)) ~ -sin(pi*x)*sin(pi*y)

# Boundary conditions
bcs = [u(0,y) ~ 0.f0, u(1,y) ~ -sin(pi*1)*sin(pi*y),
       u(x,0) ~ 0.f0, u(x,1) ~ -sin(pi*x)*sin(pi*1)]
# Space and time domains

domains = [x ∈ IntervalDomain(0.0,1.0),
           y ∈ IntervalDomain(0.0,1.0)]
pde_system = PDESystem(eq,bcs,domains,[x,y],[u(x,y)])
```

Notice that this code for specifying the partial differential equation is purely symbolic and independent of the choice of discretization, and thus at this point it can still be used for PINNs or traditional methods. To turn it into a PINN training problem, we first have to define a neural network on which to train. We can use any Flux `Chain` or DiffEqFlux `FastChain` compatible neural network definitions. This means that arbtitrary neural networks from Flux, such as RNNs, LSTMs, CNNs, etc. can all be used, such as more exotic objects like neural ODEs. Here we will use the `FastChain` with multilayer perceptrons (MLPs) for simplicity. This looks like:

```julia
# Neural network
using DiffEqFlux
dim = 2 # must match the number of input dimensions to u
chain = FastChain(FastDense(dim,16,Flux.σ),
                  FastDense(16,16,Flux.σ),
                  FastDense(16,1))
```

The equation is then "discretized" into a numerical optimization problem by defining the `PhysicsInformedNN` discretizer and passing it to the generic discretize. `PhysicsInformedNN` requires the definition of a training strategy, which here we will choose to be `GridTraining` with the same `dx` in both the x and y dimension. This looks like:

```julia
# Discretization
dx = 0.05
discretization = PhysicsInformedNN(chain,GridTraining(dx))
prob = discretize(pde_system,discretization)
```

This `prob` is now a GalacticOptim.jl `OptimizationProblem` which can be solved as any other optimization problem. To do this, we will use the `ADAM` optimizer from Flux.jl:

```julia
#Optimizer
using Flux, GalacticOptim
opt = Optim.BFGS()

#Callback function
cb = function (p,l)
    println("Current loss is: $l")
    return false
end

res = GalacticOptim.solve(prob, opt, cb = cb, maxiters=1000)
```

`res` is the result of the optimization, where `res.u` gives the optimal trained parameters, here the weights of the neural network. `discretization.phi` gives the internal representation of $u(x,y)$, and thus we can use it to visualize the solution via:

```julia
using Plots

xs,ys = [domain.domain.lower:dx/10:domain.domain.upper for domain in domains]
analytic_sol_func(x,y) = (sin(pi*x)*sin(pi*y))/(2pi^2)

u_predict = reshape([first(phi([x,y],res.minimizer)) for x in xs for y in ys],
                    (length(xs),length(ys)))
u_real = reshape([analytic_sol_func(x,y) for x in xs for y in ys],
                 (length(xs),length(ys)))
diff_u = abs.(u_predict .- u_real)

p1 = plot(xs, ys, u_real, linetype=:contourf,title = "analytic");
p2 = plot(xs, ys, u_predict, linetype=:contourf,title = "predict");
p3 = plot(xs, ys, diff_u,linetype=:contourf,title = "error");
plot(p1,p2,p3)
```

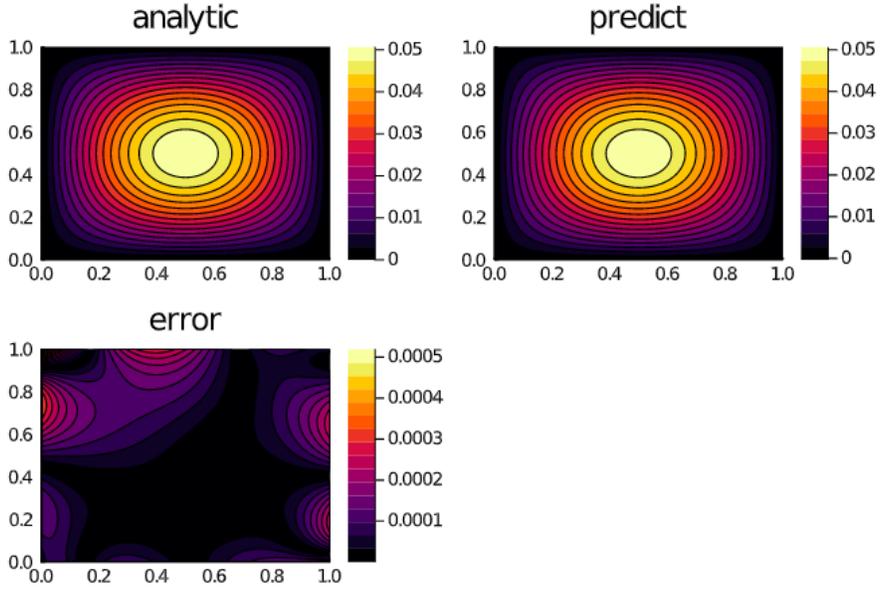

Figure 1: Plot generated by the Poisson code example.

### 3.3 Using GPUs

Neural network computations generally involve BLAS3 operations (matrix multiplications) which are heavily parallelizable by SIMD-specific hardware such as GPUs and TPUs. NeuralPDE.jl is able to make use of the generic Julia tooling for such hardware, such as CUDA.jl [10, 9], which allow for automatically transforming array-based operations to such hardware. Figure 2 demonstrates the CPU vs GPU performance benchmarks for the 2-dimensional PDE:

$$\frac{\partial u(t,x,y)}{\partial t} = \frac{\partial^2 u(t,x,y)}{\partial x^2} + \frac{\partial^2 u(t,x,y)}{\partial y^2} \tag{42}$$

with the initial and boundary conditions:

$$\begin{aligned}
u(0,x,y) &= e^{x+y}\cos(x+y), \\
u(t,0,y) &= e^y\cos(y+4t), \\
u(t,2,y) &= e^{2+y}\cos(2+y+4t), \\
u(t,x,0) &= e^x\cos(x+4t), \\
u(t,x,2) &= e^{x+2}\cos(x+2+4t).
\end{aligned} \tag{43}$$

with various number of input points and the number of neurons in the hidden layer, measuring the time for 100 iterations. In terms of code, changing the computations to the GPU simply requires moving the parameter values to the GPU, which is done via the `gpu` command. For example:

```
initθ = DiffEqFlux.initial_params(chain) |> gpu
```

Other than the `gpu` command, the setup and solving of the PDE follows exactly the format laid out in Section 3.2.

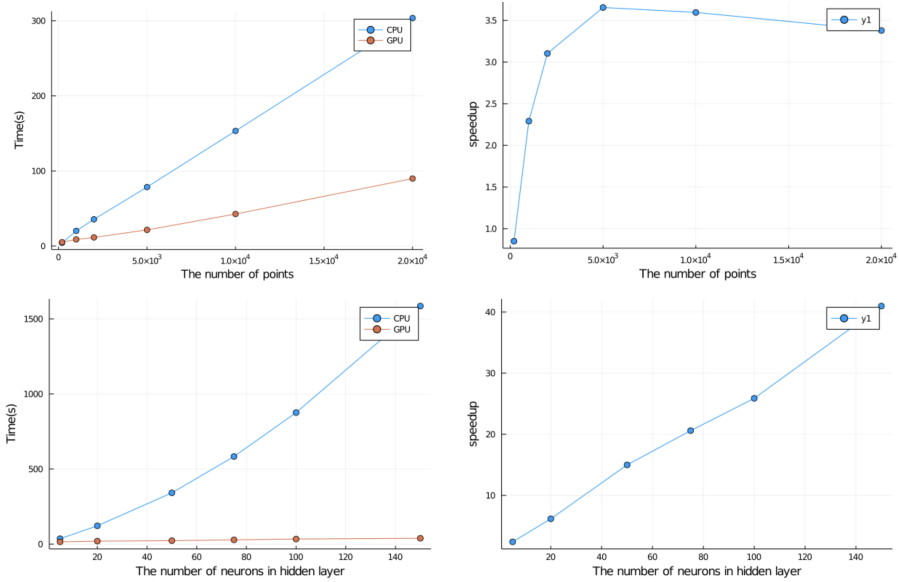

Figure 2: GPU performance across different sizes of discretizations and neural networks.

## 3.4 Systems of PDEs

As previously mentioned in Section 2.1, there are two approaches to handling systems of PDEs. One is to treat the partial differential equation as having an array-valued and thus let $N(x)$ be $\mathbb{R}^n \to \mathbb{R}^m$ for a system of $m$ equations with $n$ independent variables. One upside to this approach is that it naturally promotes larger BLAS operations, when split into multiple single-output neural networks $N_i$ one can easily run into performance issues with kernel saturation. However, there are many downsides to this approach. For one, when utilizing automatic differentiation, both forward and reverse mode automatic differentiation scale with the number of outputs, $\mathcal{O}(mn)$ and $\mathcal{O}(m+n)$ respectively, and thus if a differentiation term is only required for a single independent variable, one can easily be paying a much higher computational cost since it will require computing the derivative w.r.t. all independent variables. This issue compounds as the derivative order increases, as this multiplicatively increases the output size, and thus will add more computational burden to PINNs with second order derivatives like diffusion terms. In addition, this approach requires that all of the dependent variables are defined on the full domain, which does not hold in PDEs like the DFN in Section 6.1. Likewise, the splitting approach can avoid kernel saturation issues by combination calculations at multiple points in the domain, alleviating its main issue. Therefore, NeuralPDE.jl uses the split network approach.

The only difference with the above examples is that, instead of supplying a single `FastChain` or `Chain` object, a system of equations requires declaring an array of dependent variables along with an array of neural networks. The input dimension of each neural network should match the dependent variables of a given independent variable.

As an example, take the partial differential-algebraic equation system (PDAE) from before [35]:

$$\frac{\partial u_1}{\partial t} = \frac{\partial^2 u_1}{\partial x^2} + u_3 \sin(\pi x) \tag{44}$$

$$\frac{\partial u_2}{\partial t} = \frac{\partial^2 u_2}{\partial x^2} + u_3 \cos(\pi x) \tag{45}$$

$$0 = u_1 \sin(\pi x) + u_2 \cos(\pi x) - e^{-t} \tag{46}$$

with initial conditions:

$$u_1(0, x) = \sin(\pi x) \tag{47}$$

$$u_2(0, x) = \cos(\pi x) \tag{48}$$

$$\frac{\partial u_1(0, x)}{\partial t} = -\sin(\pi x) \tag{49}$$

$$\frac{\partial u_2(0, x)}{\partial t} = -\cos(\pi x) \tag{50}$$

$$\tag{51}$$

We would represent this using the following `PDESystem`:

```
using ModelingToolkit

@parameters t, x
@variables u1(..), u2(..), u3(..)
Dt = Differential(t)
Dtt = Differential(t)^2
Dx = Differential(x)
Dxx = Differential(x)^2

eqs = [Dtt(u1(t,x)) ~ Dxx(u1(t,x)) + u3(t,x)*sin(pi*x),
       Dtt(u2(t,x)) ~ Dxx(u2(t,x)) + u3(t,x)*cos(pi*x),
       0. ~ u1(t,x)*sin(pi*x) + u2(t,x)*cos(pi*x) - exp(-t)]

bcs = [u1(0,x) ~ sin(pi*x),
       u2(0,x) ~ cos(pi*x),
       Dt(u1(0,x)) ~ -sin(pi*x),
```

```julia
        Dt(u2(0,x)) ~ -cos(pi*x),
        u1(t,0) ~ 0.,
        u2(t,0) ~ exp(-t),
        u1(t,1) ~ 0.,
        u2(t,1) ~ -exp(-t)]

# Space and time domains
domains = [t ∈ IntervalDomain(0.0,1.0),
           x ∈ IntervalDomain(0.0,1.0)]
pde_system = PDESystem(eqs,bcs,domains,[t,x],[u1(t,x),u2(t,x),u3(t,x)])
```

Now we design our discretization to be performed by an array of 3 neural networks:

```julia
# Neural network
input_ = length(domains)
n = 20
chain1 = FastChain(FastDense(input_,n,Flux.σ),FastDense(n,n,Flux.σ),FastDense(n,1))
chain2 = FastChain(FastDense(input_,n,Flux.σ),FastDense(n,n,Flux.σ),FastDense(n,1))
chain3 = FastChain(FastDense(input_,n,Flux.σ),FastDense(n,n,Flux.σ),FastDense(n,1))

strategy = QuadratureTraining()
discretization = PhysicsInformedNN([chain1,chain2,chain3], strategy)
```

Now we solve the PDE:

```julia
prob = discretize(pde_system,discretization)

cb = function (p,l)
    println("Current loss is: $l")
    return false
end

res = GalacticOptim.solve(prob,BFGS(); cb = cb, maxiters=200)
prob = remake(prob,u0=res.minimizer)
res = GalacticOptim.solve(prob,ADAM(10^-2); cb = cb, maxiters=10000)
prob = remake(prob,u0=res.minimizer)
res = GalacticOptim.solve(prob,BFGS(); cb = cb, maxiters=200)
phi = discretization.phi
```

and plot the result:

```julia
using Plots

ts,xs = [domain.domain.lower:0.01:domain.domain.upper for domain in domains]

initθ = discretization.init_params
```

```
acum = [0;accumulate(+, length.(initθ))]
sep = [acum[i]+1 : acum[i+1] for i in 1:length(acum)-1]
minimizers = [res.minimizer[s] for s in sep]

analytic_sol_func(t,x) = [exp(-t)*sin(pi*x), exp(-t)*cos(pi*x), (1+pi^2)*exp(-t)]
u_real  = [[analytic_sol_func(t,x)[i] for t in ts for x in xs] for i in 1:3]
u_predict  = [[phi[i]([t,x],minimizers[i])[1] for t in ts  for x in xs] for i in 1:3]
diff_u = [abs.(u_real[i] .- u_predict[i] ) for i in 1:3]

for i in 1:3
    p1 = plot(ts, xs, u_real[i], st=:surface,title = "u$i, analytic");
    p2 = plot(ts, xs, u_predict[i], st=:surface,title = "predict");
    p3 = plot(ts, xs, diff_u[i],linetype=:contourf,title = "error");
    plot(p1,p2,p3)
    savefig("sol_u$i")
end
```

In Figure 3 we show the result of the first $u_1$.

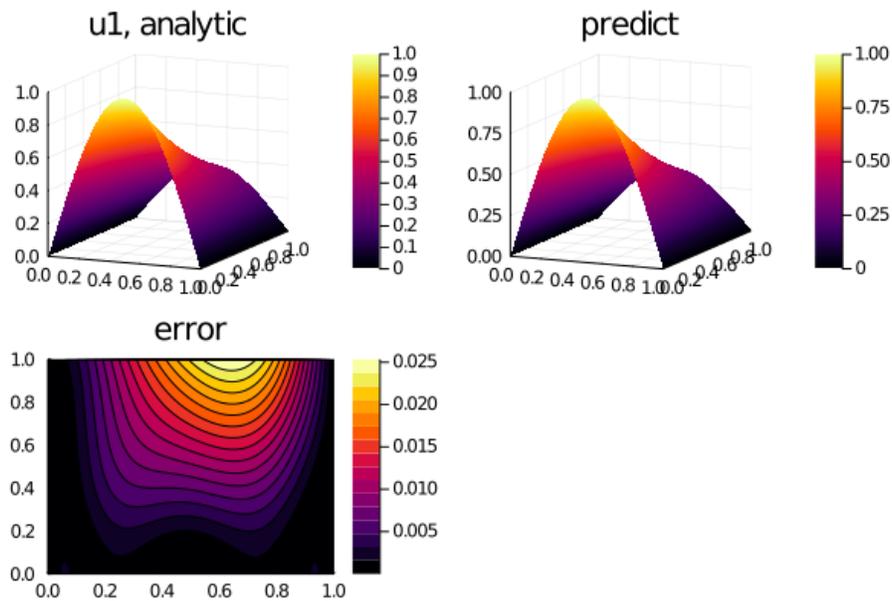

Figure 3: Plot generated by the Systems of Equations code example.

## 3.5 Lorenz Parameter Estimation Example (Inverse Problems)

We define the parameter estimation problem of Lorenz Systems as,

$$\frac{dx}{dt} = \sigma(y - x), \tag{52}$$

$$\frac{dy}{dt} = x(\rho - z), \tag{53}$$

$$\frac{dz}{dt} = xy - \beta z \tag{54}$$

and the boundary conditions,

$$x(0) = 1, \tag{55}$$
$$y(0) = 0, \tag{56}$$
$$z(0) = 0 \tag{57}$$

which corresponds to the following PDE definition code:

```julia
@parameters t ,σ ,β, ρ
@variables x(..), y(..), z(..)
Dt = Differential(t)
eqs = [Dt(x(t)) ~ σ*(y(t) - x(t)),
       Dt(y(t)) ~ x(t)*(ρ - z(t)) - y(t),
       Dt(z(t)) ~ x(t)*y(t) - β*z(t)]

bcs = [x(0) ~ 1.0, y(0) ~ 0.0, z(0) ~ 0.0]
domains = [t ∈ IntervalDomain(0.0,1.0)]
pde_system = PDESystem(eqs,bcs,domains,[t],[x(t), y(t), z(t)],[σ, ρ, β],
                    defaults=Dict([p .=> 1.0 for p in [σ, ρ, β]]))
```

We will optimize the parameters $\sigma$, $\beta$ and $\rho$ according to the available data points. We add an additional loss function based that calculates the loss between the trial solution and available data points.

```julia
function additional_loss(phi, θ , p)
    return sum(sum(abs2, phi[i](t_ , θ[sep[i]]) .- u_[[i], :])/len for i in 1:1:3)
end
```

Notice that 'u_' and 't_' are vectors containing the data points of solutions (x , y , z) at times t. We define the neural networks for x , y and z using the `FastChain`. We will use shallow neural networks to represent the solution:

```julia
using DiffEqFlux
input_ = length(domains)
n = 8
chain1 = FastChain(FastDense(input_,n,Flux.σ),FastDense(n,n,Flux.σ),
                   FastDense(n,n,Flux.σ),FastDense(n,1))
```

```
chain2 = FastChain(FastDense(input_,n,Flux.σ),FastDense(n,n,Flux.σ),
                    FastDense(n,n,Flux.σ),FastDense(n,1))
chain3 = FastChain(FastDense(input_,n,Flux.σ),FastDense(n,n,Flux.σ),
                    FastDense(n,n,Flux.σ),FastDense(n,1))
```

And using the `PhysicsInformedNN` interface we discretize the problem.

```
dt = 0.01
discretization = NeuralPDE.PhysicsInformedNN([chain1 , chain2, chain3],
                    NeuralPDE.GridTraining(dt),param_estim=true,
                    additional_loss=additional_loss)
prob = NeuralPDE.discretize(pde_system,discretization)
```

Here while defining the discretization we pass the argument `param_estim=` true, which allows us to use the parameter values within the extended loss (see Section 2.6 which describes how the extended loss requires using the $\lambda$ parameters to solve the inverse problem). Along with this, we pass the `additional_loss` function, which is added to the total loss function. Notice that while defining the `PDESystem` we pass an additional vector for parameters and a dictionary for initial values of the parameters.

And then we solve the optimization problem,

```
cb = function (p,l)
    println("Current loss is: $l")
    return false
end

res = GalacticOptim.solve(prob, BFGS(); cb = cb, maxiters=5000)
p_ = res.minimizer[end-2:end] # p_ = [9.93, 28.002, 2.667]
```

As mentioned in earlier sections, `res` is the result of the optimizer, the resulting parameters are the last three elements of `res.minimizer`. Since the data was generated from the Lorenz equations with $[10, 28, 2.66]$ as the parameters, this demonstrates the successful solution of the inverse problem by a PINN.

# 4 Numerical Experiments of PINN Training Techniques

This section is devoted to the comparison between different training strategies and minimizers on various classes of differential equations.
To that end, all the examples are run with a modified callback function, which evaluates the error on the same grid for every strategy: the algorithm *GridTraining(dx)* with a mesh size of $dx = 0.1$ is used. This way we are sure to conduct an unbiased comparison, since we are removing the possible discrepancies due to the specific loss function computed by each strategy internally.

The four Quadrature-based strategies considered are the following:

| Quadrature Alg | reltol | abstol | maxiters |
|---|---|---|---|
| CubaCuhre | 1 | $1 \times 10^{-4}$ | 100 |
| HCubatureJL | 1 | $1 \times 10^{-4}$ | 100 |
| CubatureJLh | 1 | $1 \times 10^{-4}$ | 100 |
| CubatureJLp | 1 | $1 \times 10^{-4}$ | 100 |

The only exception is the Diffusion example, for which $abstol = 10^{-5}$ was used.

These evaluate the loss function by approximating the integral of the PDE component of the loss function on the training domain through adaptive quadrature methods. 4 lists the specific Quadrature algorithms used, with the chosen relative and absolute tolerance and the fixed number of iterations.
In addition, the comparison extends to:

- *GridTraining(0.1)*: the points are initialized on an equally spaced mesh of step 0.1 on a hypercube of the same dimension as the training domain. The sampling points are never the same used for the error evaluation (for the Diffusion example, *GridTraining([0.2,0.1])* is employed);

- *StochasticTraining(100)* : 100 points sampled from the domain are stochastically selected at each iteration;

- *QuasiRandomTraining(100; sampling_alg = UniformSample(), minibatch = 100)* : 100 points for every minibatch are generated with a quasi-Monte Carlo sampling algorithm.

The main minimizer used is ADAM (Adaptive Moment Estimation), a method of stochastic gradient which computes an adaptive learning by taking into account the first moments of the gradient. In some examples it is compared with L-BFGS (a limited memory variant of the standard Broyden–Fletcher–Goldfarb–Shanno algorithm), which computes gradient descent from the approximation of the Hessian Matrix. Even though L-BFGS converges faster, i.e. reaches a fixed error with less iterations, ADAM proves to be faster time-wise, i.e. performs more iterations in a shorter amount of time, especially for high dimensional problems. In addition, L-BFGS often stops before completing the fixed amount of iterations.

## 4.1   Example 1: 1-D Diffusion

As a simple example, let us consider the one-dimensional Diffusion equation which describes several physical scenarios such as the heat conduction in a one-dimensional solid body, spread of a die in a stationary fluid, population dispersion and similar phenomena.

The simple diffusion equation can be derived by considering a set of one-dimensional random walkers in the continuum limit. Let us assume that our walkers move on a one-dimensional lattice and at every $\Delta t$, choose randomly with probability $\frac{1}{2}$, to move rightward or leftward, with a step size $\Delta l$. Let us consider $n(x,t)$, the number of walkers at position $x$ at time $t$, its change in time is

$$n(x, t + \Delta t) - n(x,t) = \frac{1}{2} n(x - \Delta l, t) + \frac{1}{2} n(x + \Delta l, t) - n(x,t) \qquad (58)$$

In the limit of small time step $\Delta t$ and displacement $\Delta l$:

$$\frac{\partial n}{\partial t} \Delta t = \frac{1}{2}\left[ n(x,t) - \Delta l \frac{\partial n}{\partial x} + \frac{1}{2}(\Delta l)^2 \frac{\partial^2 n}{\partial x^2}\right] +$$
$$+ \frac{1}{2}\left[ n(x,t) + \Delta l \frac{\partial n}{\partial x} + \frac{1}{2}(\Delta l)^2 \frac{\partial^2 n}{\partial x^2}\right] - n(x,t) = \frac{1}{2}(\Delta l)^2 \frac{\partial^2 n}{\partial x^2} \qquad (59)$$

which gives the diffusion equation

$$\frac{\partial n}{\partial t} = D \frac{\partial^2 n}{\partial x^2} \qquad (60)$$

where $D = \frac{(\Delta l)^2}{2\Delta t}$ is the diffusion coefficient and n(x,t) is now a density of walkers, that is $n(x,t)dx$ is the number of walkers in the interval $[x, x + dx]$ at time $t$. The same result can be obtained by viewing the diffusion equation as a consequence of the continuity equation and Fick's $1^{st}$ law, hence the name of Fick's $2^{nd}$ law [16].

In our case, by leaving aside the random walk analogy and considering a general function $u(x,t)$, we are going to focus on the following specific equation, in the presence of an additional potential:

$$\frac{\partial u(t,x)}{\partial t} - D \frac{\partial^2 u(t,x)}{\partial x^2} = f(x,t) \qquad (61)$$

with Dirichlet boundary condition:

$$\begin{aligned}
&f(x,t) = (e^{-t} - \pi^2) \cdot \sin(\pi x) \\
&u(x,0) = \sin(\pi x),\ x \in [-1,1],\ t \in [0,1] \\
&u(-1,t) = 0, \\
&u(1,t) = 0.
\end{aligned} \qquad (62)$$

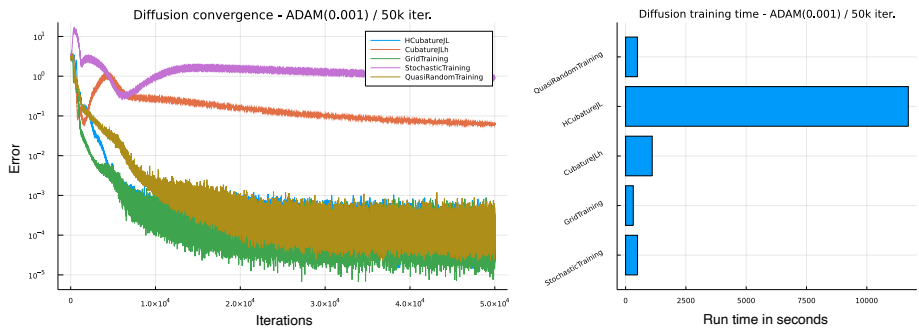

Figure 4: Convergence of the diffusion equation training using ADAM as minimizer.

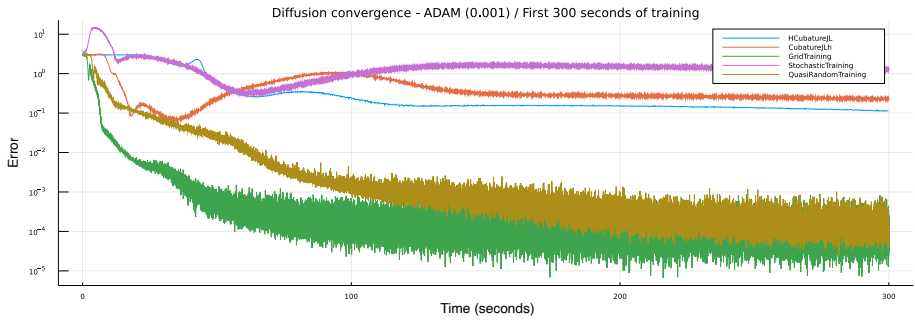

Figure 5: Training convergence from 0 to 300 seconds.

| Strategy | Minimum Error Value |
| --- | --- |
| HCubatureJL | $1.41 \times 10^{-5}$ |
| CubatureJLh | $5.68 \times 10^{-2}$ |
| GridTraining | $6.60 \times 10^{-6}$ |
| StochasticTraining | $3.06 \times 10^{-1}$ |
| QuasiRandomTraining | $2.06 \times 10^{-5}$ |

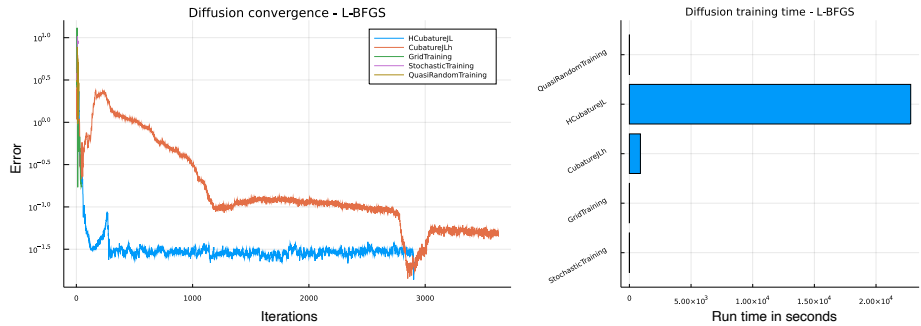

Figure 6: Convergence of the diffusion equation training using L-BFGS as minimizer.

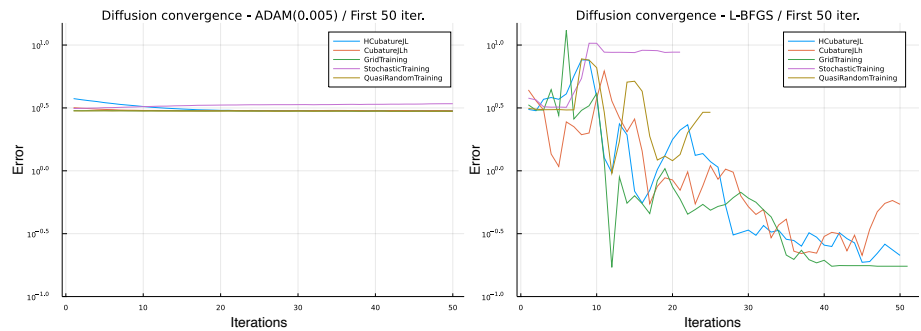

Figure 7: Convergence speed comparison between ADAM and L-BFGS.

The following plots show the discrepancy between the prediction of the Physics-Informed Neural Network and the analytical solution to the equation. This proves that the quadrature strategy HCubatureJL, as well as GridTraining and QuasiRandomTraining, not only reach a low loss value, but converge to the correct solution. The remaining two strategies do not perform as well in this case.

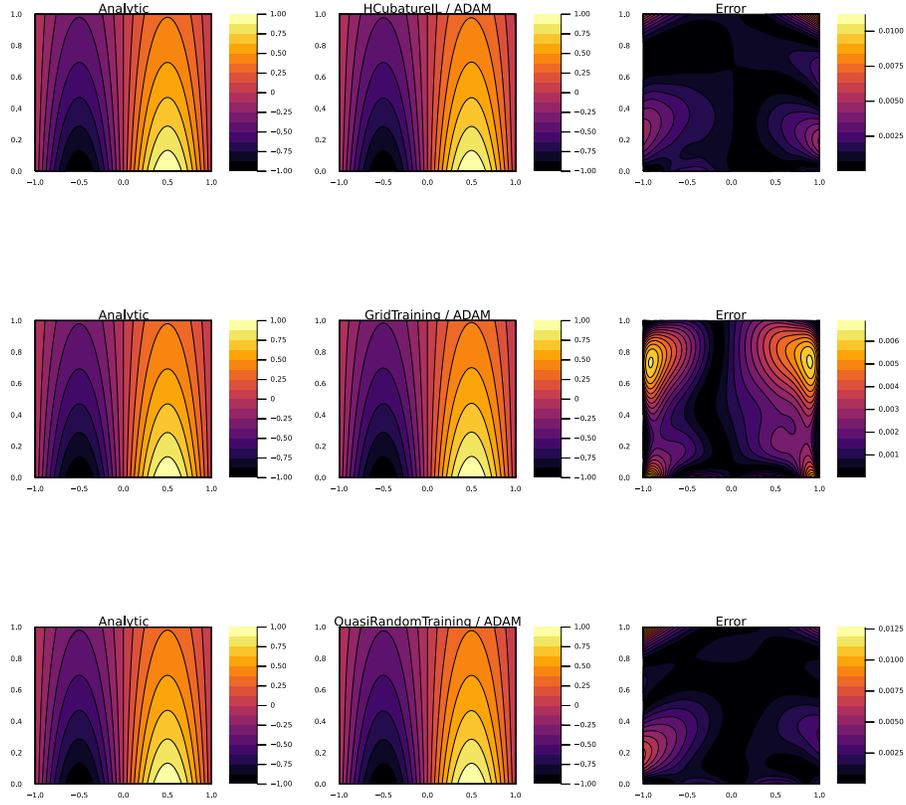

Figure 8: Comparison of the PINN's prediction with the analytical solution.

## 4.2 Example 2: Poisson-Nernst-Planck

The following example presents a model depicting the ion transport of $Cl^-$ and $Na^+$, solely governed by diffusion and migration, and described by the Poisson–Nernst–Planck equations. The model is a simplification based on [56, 32], where an electrochemical treatment was used for tumor ablation. The equations are written as:

$$\frac{\partial C_B}{\partial t} + \nabla \cdot \mathbf{N_B} = \mathbf{R_B} \tag{63}$$

Here, the molar flux is:

$$\mathbf{N_B} = -D_B \mathbf{C_B} - \frac{\mathbf{z_B}}{|\mathbf{z_B}|} \mathbf{u_B} \mathbf{C_B} \tag{64}$$

where $B = \{Cl^-, Na^+\}$, $D_B$, $C_B$, $z_B$ and $u_B$ are the chemical species, diffusion coefficient, concentration, charge number (with sign) and mobility of the species $B$, $\Phi$ is the electric potential and $t$ is the time.

The ionic mobility and diffusion coefficient are related through the Nernst-Einstein equation:

$$D_B = \frac{RT}{|z_B|F}u_B \tag{65}$$

where $T$ is the absolute temperature, $R$ the universal gas constant and $F$ the Faraday's constant.

$R_B$ represents the production of species $B$ through chemical reactions in the electrolyte.

$$R_{Cl^-} = R_{Na^+} = 0 \tag{66}$$

Combining equations 61, 64, 65, and 66

$$\frac{\partial C_{Cl^-}}{\partial t} = \nabla \cdot (D_{Cl^-}\nabla C_{Cl^-} + \frac{z_{Cl^-}}{|z_{Cl^-}|}u_{Cl^-}C_{Cl^-}\nabla\Phi) \tag{67}$$

$$\frac{\partial C_{Na^+}}{\partial t} = \nabla \cdot (D_{Na^+}\nabla C_{Na^+} + \frac{z_{Na^+}}{|z_{Na^+}|}u_{Na^+}C_{Na^+}\nabla\Phi) \tag{68}$$

If there is no applied magnetic field and current-induced magnetic fields are neglected as well as those induced by time-varying electric fields, then the electrodynamic problem boils down to an electrostatic problem. This problem can be specified by the Poisson equation, which relates the spatial variation of the electric field with the distribution of charges, which, for a medium with constant dielectric constant, can be expressed as

$$\nabla^2\Phi = -\frac{F}{\epsilon}\sum_B z_B \, C_B \tag{69}$$

Table 4.2 shows the electrochemical parameter values of the present model.

Initial $Na^+$ and $Cl^-$ concentrations are set to 0.16 M on the entire domain, while initial electric potential is set to zero. Boundary conditions based on the Butler-Volmer equations were simplified to Dirichlet boundary conditions. At the anode, $Na^+$ and $Cl^-$ concentrations are set to 0 and 0.22 $mol\ dm^{-3}$, respectively, while at the cathode, these concentrations are set to 0.32 and 0 $mol\ dm^{-3}$, respectively. The potential difference between anode and cathode is 4V.

To simplify the analysis of the physical parameters, a nondimensionalizationn algebraic transformation was performed. Table 4.2 shows reference parameters for time, distance, concentration, and potential, as well as the nondimensional Péclet, Migration, and Poisson numbers.

The equation system was solved using NeuralPDE, and the code is freely available at `https://github.com/emmanuellujan/1d-poission-nernst-planck`.

| Parameter | Value | Unit |
|---|---|---|
| $t_{max}$ | 1.0 | $s$ |
| $x_{max}$ | 0.38 | $dm$ |
| $D_{Na}$ | 0.89e-7 | $dm^2\ s^{-1}$ |
| $D_{Cl}$ | 1.36e-7 | $dm^2\ s^{-1}$ |
| $z_{Na}$ | 1.0 | $m^2\ V^{-1}\ s^{-1}$ |
| $z_{Cl}$ | -1.0 | $m^2\ V^{-1}\ s^{-1}$ |
| $Na_0$ | 0.16 | $mol\ dm^{-3}$ |
| $Cl_0$ | 0.16 | $mol\ dm^{-3}$ |
| $\Phi_0$ | 4.0 | $V$ |
| $Na_{anode}$ | 0.0 | $mol\ dm^{-3}$ |
| $Na_{cathode}$ | 0.32 | $mol\ dm^{-3}$ |
| $Cl_{anode}$ | 0.22 | $mol\ dm^{-3}$ |
| $Cl_{cathode}$ | 0.0 | $mol\ dm^{-3}$ |
| $F$ | 96485.3415 | $A\ s\ mol^{-1}$ |
| $R$ | 831.0 | $kg\ dm^2\ K^{-1}\ mol^{-1}\ s^{-2}$ |
| $T$ | 298.0 | $K$ |
| $\epsilon$ | 78.5 | $K$ |

Table 1: Electrochemical parameters of Poisson-Nernst-Planck example.

| Parameter | Value | Unit | Parameter | Value | Unit |
|---|---|---|---|---|---|
| $t_{ref}$ | 1.0 | $s$ | $Pe_B$ | $\frac{x_{ref}^2}{t_{ref}*D_B}$ | |
| $x_{ref}$ | 0.38 | $dm$ | $M_B$ | $\frac{x_{ref}^2}{t_{ref}*\Phi_{ref}*u_B}$ | |
| $C_{ref}$ | 0.16 | $mol\ dm^{-3}$ | $Po$ | $\frac{\epsilon*Phi_{ref}}{F*x_{ref}*C_{ref}}$ | |
| $\Phi_{ref}$ | 1.0 | $V$ | | | |

Table 2: Parameters for non-dimensional analysis of Poisson-Nernst-Planck example.

### 4.3 Example 3: Burgers' equation

This case study analyzes Burgers' equation, a simplified version of the Navier-Stokes equations [7]. It is written as follows:

$$\frac{\partial u(t,x)}{\partial t} + u(t,x)\frac{\partial u(t,x)}{\partial x} = \nu\frac{\partial^2 u(t,x)}{\partial x^2} \tag{70}$$

where $u$ is a scalar field, $\nu$ is the diffusion coefficient or kinematic viscosity, $x$ is the spatial variable, and $t$ is the time. The initial condition for this problem

is:

$$u(0, x) = -\frac{2\nu}{\phi(x)} \frac{\partial \phi(x)}{\partial x} + 4 \tag{71}$$

$$\phi(x) = \exp\left(\frac{-x^2}{4\nu}\right) + \exp\left(\frac{-(x - 2\pi)^2}{4\nu}\right) \tag{72}$$

The periodic boundary condition is:

$$u(t, 0) = u(t, 2\pi) \tag{73}$$

The analytical solution of this problem is given by:

$$u(t, x) = -\frac{2\nu}{\phi} \frac{\partial \phi(t, x)}{\partial x} + 4 \tag{74}$$

$$\phi(t, x) = \exp\left(\frac{-(x - 4t)^2}{4\nu(t + 1)}\right) + \exp\left(\frac{-(x - 4t - 2\pi)^2}{4\nu(t + 1)}\right) \tag{75}$$

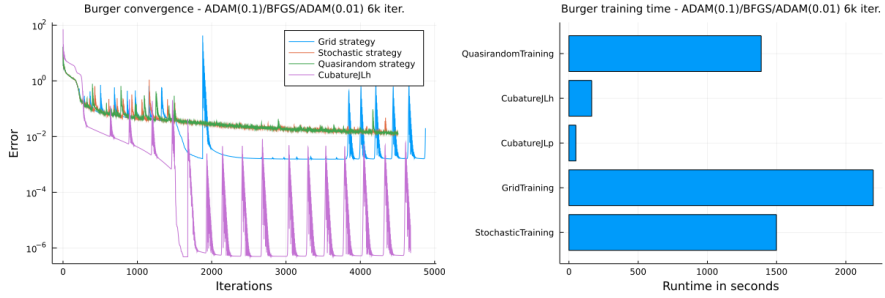

Figure 9: Convergence of the Burgers' equation training using ADAM and BFGS as minimizers.

| Strategy | Minimum Error Value |
|---|---|
| QuasiRandomTraining | $1.28 \times 10^{-2}$ |
| CubatureJLh | $6.6 \times 10^{-7}$ |
| CubatureJLp | $1.13 \times 10^{-18}$ |
| GridTraining | $2.0 \times 10^{-2}$ |
| StochasticTraining | $1.19 \times 10^{-2}$ |

https://github.com/emmanuellujan/burgers-equation-neuralpde

### 4.4 Example 4: Level-set equation

The level-set equation is a non-linear partial differential equation at the core of various models in image processing, computational biology, computational fluid dynamics, climate physics and environmental physics. One of the advantages of the level-set method is that it makes possible numerical computations on surfaces and curves without having to parametrize these objects. As an example application, it can be used to predict the spread of wildfires. Let us consider a fire burning in an area $\Omega = \Omega(t)$ and let's project the terrain on the $(x, y)$ plane. The burning area can be studied by introducing the level set function $\psi = \psi(x, t)$ where the burning area at time $t$ is:

$$\Omega(t) = \{x : \psi(x, y, t) \leq 0\} \qquad (76)$$

The boundary of this region, described by the contour surface:

$$\Gamma(t) = \{x : \psi(x, y, t) = 0\}, \qquad (77)$$

represents the fireline, while the region where $\psi(x, y, t) > 0$ hasn't encountered the fire yet. The level set function presented above can be retrieved as a solution of the following partial differential equation

$$\partial_t \psi + S ||\nabla \psi|| = 0, \qquad (78)$$

called the *level set equation*, where the fire spread rate $S$ is computed from fuel properties, using the modified Rothermel formula [47]:

$$S = R_0(1 + \phi_W + \phi_S). \qquad (79)$$

Here, $R_0$ represents the spread rate in the absence of wind, whereas $\phi_W$ and $\phi_S$ are respectively the wind factor and the slope factor.
On the fireline, from Eq.(77), the tangential component of the gradient $\nabla \psi$ is zero, meaning that the fire propagation speed is normal to the fireline $\Gamma$: the model postulates $S$ as a function of the normal component of the wind factor $U$ and the terrain gradient $\nabla z$ [47].
Some common boundary conditions for such model can be, for example, a cone or a paraboloid with a non-void intersection with the plane $z = 0$ at time $t = 0$, representing the kernel of a wildfire at the initial state. In our example the initial condition is set to

$$\psi(x, y, 0) = \sqrt{(x - x_{ignition})^2 + (y - y_{ignition})^2} - A \qquad (80)$$

and the values for the parameters mentioned above are the following:
    The architecture is a feed-forward NN with one input layer of 3 neurons (two spatial and one temporal dimension), one hidden layer of 16 neurons and an output layer of dimension one. The input and the hidden layer have a sigmoid activation function, the output layer has a linear output.
After 20k iterations using ADAM with an initial learning rate of 0.005 we get the following training plots:

| Variable | Value | Unit |
|---|---|---|
| $R_0$ | 0.1125 | $m/s$ |
| $U_{wind}$ | $[0.0, 2.0]$ | |
| $\phi_w$ | $0.157 max((0.44\langle\frac{\nabla u}{\|\nabla u\|}, U_{wind}\rangle)^{0.041}, 1.45)$ | |
| $\phi_s$ | 0 | |
| $x_{ignition}$ | 0.5 | $m$ |
| $y_{ignition}$ | 0.5 | $m$ |
| $A$ | 0.2 | $m$ |

Table 3: Parameters of level-set example.

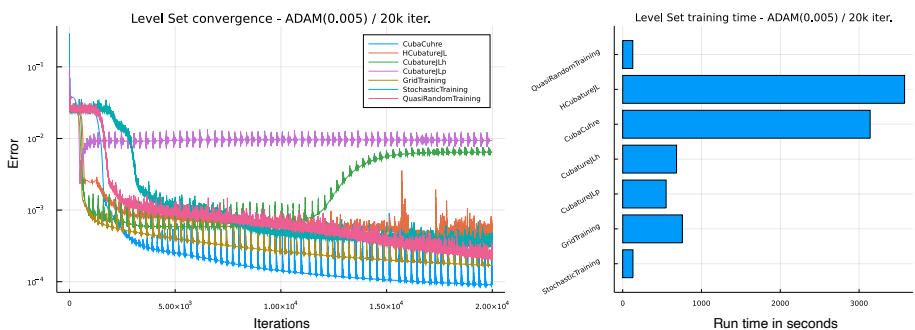

Figure 10: Convergence of the Level Set equation training using ADAM as minimizer.

The comparison shows that for low-dimensional equations Quadrature-based strategies are not always significantly better than the standard Grid training: the latter reaches a minimum error value which is slightly higher that CubaCuhre's one, but is significantly faster in this case. However, the Quadrature approach is able to achieve a relatively accurate loss. As expected, Stochastic and QuasiRandom strategies are the fastest, but suffer from more frequent bumps during the training due to the inherently stochastic nature of the algorithms. Follows a more precise list of the final results:

| Strategy | Minimum Error Value |
|---|---|
| CubaCuhre | $9.05 \times 10^{-5}$ |
| HCubatureJL | $3.72 \times 10^{-4}$ |
| CubatureJLh | $5.52 \times 10^{-4}$ |
| CubatureJLp | $2.34 \times 10^{-3}$ |
| GridTraining | $1.68 \times 10^{-4}$ |
| StochasticTraining | $3.00 \times 10^{-4}$ |
| QuasiRandomTraining | $2.05 \times 10^{-4}$ |

The following plot shows the trend of the loss as a function of time, and provides new information: for example, it is possible to notice that QuasiRandom is the fastest strategy convergence-wise, while CubaCuhre decreases more gradually, even though the latter reaches a lower error value in the end.

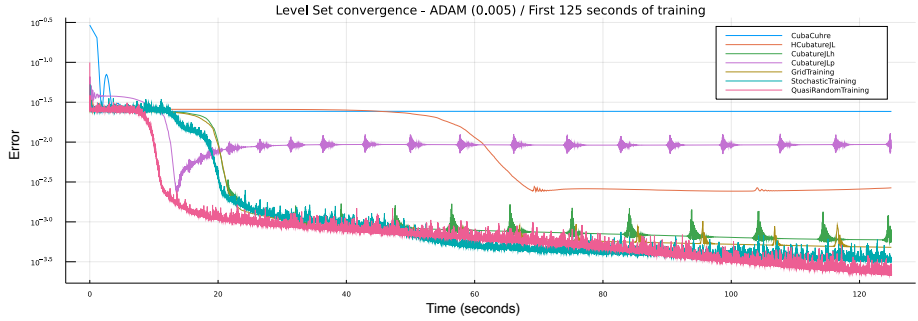

Figure 11: Training convergence with time on the horizontal axis (first 220 seconds of training).

In addition, we provide the corresponding plots obtained using L-BFGS as a minimizer. The comparison is two-fold. On one hand, Fig.12 proves that L-BFGS is much slower than ADAM time-wise, especially with CubaCuhre and CubatureJLh. Furthermore, as pointed out in Section 4, sometimes it stops after few iterations, especially with Stochastic and QuasiRandom strategies, because the calculated gradient approaches zero within the predetermined threshold.

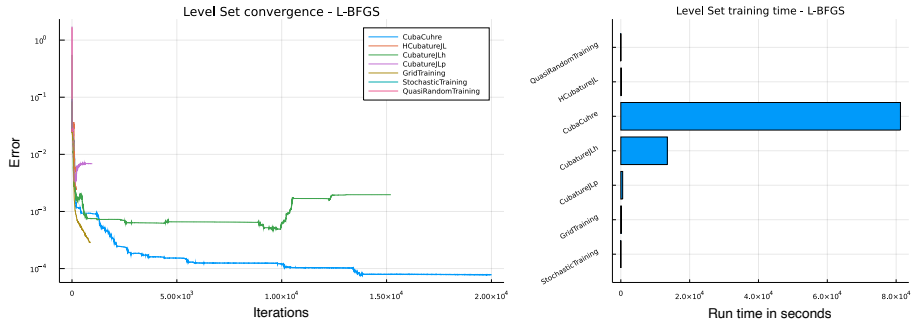

Figure 12: Convergence of the Level Set equation training using L-BFGS as minimizer.

The minimum error values obtained are the following:

| Strategy | Minimum Error Value |
|---|---|
| CubaCuhre | $7.71 \times 10^{-5}$ |
| HCubatureJL | $2.39 \times 10^{-3}$ |
| CubatureJLh | $4.87 \times 10^{-4}$ |
| CubatureJLp | $3.28 \times 10^{-3}$ |
| GridTraining | $2.86 \times 10^{-4}$ |
| StochasticTraining | $2.87 \times 10^{-2}$ |
| QuasiRandomTraining | $2.46 \times 10^{-2}$ |

On the other hand, Fig.13 shows that L-BFGS is faster convergence-wise. After 200 iterations, it reaches an error of approximately one order of magnitude less than ADAM. In conclusion, ADAM proves to be more widely applicable and more efficient for practical use.

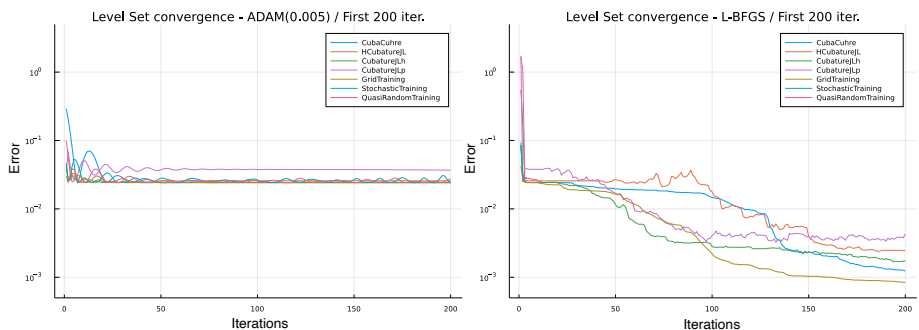

Figure 13: Convergence speed comparison between ADAM and L-BFGS.

## 4.5 Example 5: Allen Cahn equation

The Allen Cahn equation is a reaction-diffusion equation that often arises in mathematical physics and models phase separation processes in multi-component alloy systems, such as order-disorder transitions, where the volume fractions of the phases can change. One of its most common applications is the the study of grain growth in crystalline samples [14]. However, it has turned out to be efficient also in image segmentation [6], motion by mean curvature flows [15] and tumor growth [48].

Most generally, it is a second-order semilinear parabolic partial differential equation of the form ([4]):

$$\frac{\partial u}{\partial t}(t,x) = \Delta u(t,x) - \epsilon^{-2} f(u), \quad u(0) = u_0, \quad \frac{\partial u}{\partial n}(t,\cdot)|_{\partial \Omega}, \qquad (81)$$

where $\epsilon$ is the width of the diffuse interface, i.e. the region of separation between the phases and $f$ is the derivative of a double well potential $F$ which satisfies $F(\pm 1) = 0$: if we take $F(s) = (s^2 - 1)/4$, then $f(s) = s^3 - s$.

Here we consider $\epsilon = 1$ and focus on the equation of the form ([21]):

$$\frac{\partial u}{\partial t}(t, x) = \Delta u(t, x) + u(t, x) - u(t, x)^3, \tag{82}$$

where $x = (x_1, x_2, x_3, x_4)$ is a four dimensional vector, which, together with the time variable $t$, constitutes the domain $[0, T] \times \Gamma$.
As in [21], we are going to use the following initial condition:

$$u(0, x) = \frac{1}{2 + 0.4 \|x\|^2}, \qquad u_0 \in L^2(\Omega). \tag{83}$$

In this case our model consists of a feed-forward NN (as in the previous case) with one input layer of 5 neurons, one hidden layer of 20 neurons and an output layer of dimension one. The input and the hidden layer have a sigmoid activation function, the output layer have a linear output.

After training for 2500 iterations using ADAM with an initial training learning rate of 0.01 we obtain these training plots:

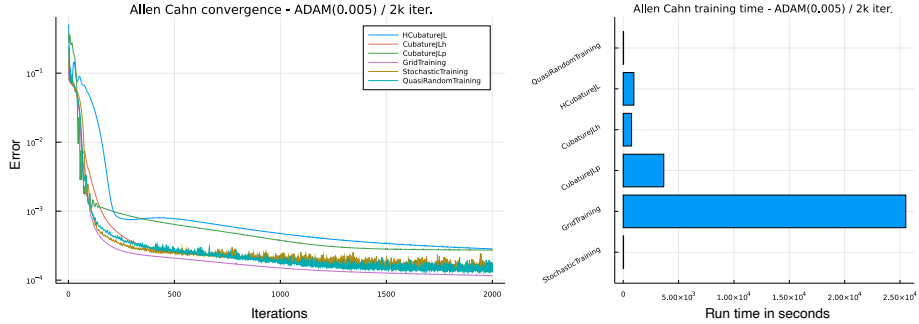

Figure 14: Convergence of the Allen Cahn equation training using ADAM as minimizer.

As expected, in this case GridTraining is the slowest by far, due to the high dimensionality of the equation, even though it is the best in terms of loss performance. QuasiRandom and StochasticTraining perform well on the time benchmark in comparison with the other strategies; in addition, Fig. 15 evidently shows that they display a steeper descent compared to the other strategies. The Quadrature strategies achieve an accurate value of the loss (as Tab. 4.5 shows) as well and prove to be much faster than GridTraining.

| Strategy | Minimum Error Value |
|---|---|
| HCubatureJL | $2.83 \times 10^{-4}$ |
| CubatureJLh | $1.46 \times 10^{-4}$ |
| CubatureJLp | $2.73 \times 10^{-4}$ |
| GridTraining | $1.16 \times 10^{-4}$ |
| StochasticTraining | $1.41 \times 10^{-4}$ |
| QuasiRandomTraining | $1.30 \times 10^{-4}$ |

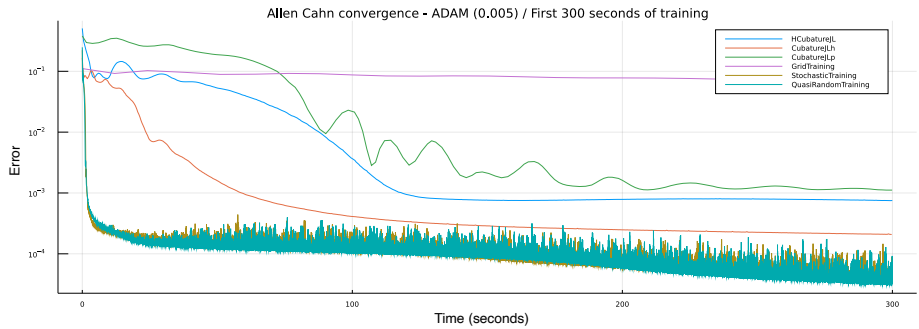

Figure 15: Training convergence with time on the horizontal axis (first 300 seconds of training).

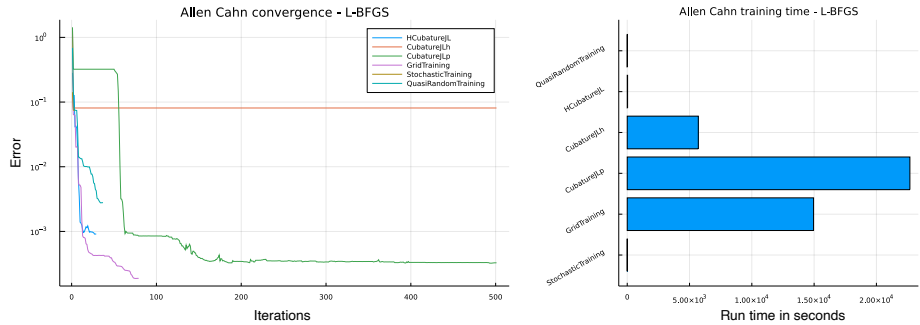

Figure 16: Convergence of the Allen Cahn equation training using L-BFGS as minimizer.

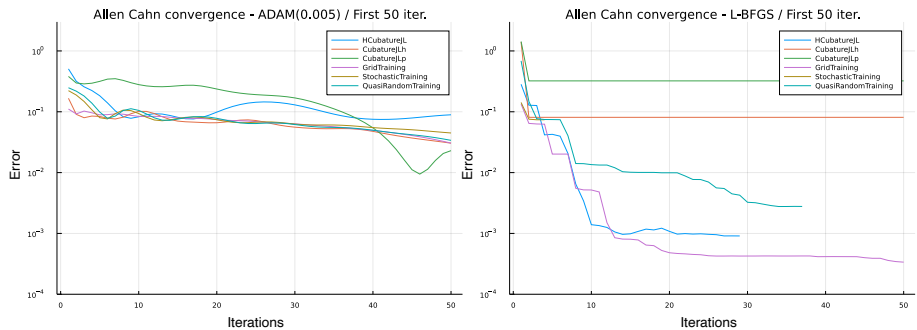

Figure 17: Convergence speed comparison between ADAM and L-BFGS.

## 4.6 Example 6: Hamilton-Jacobi-Bellman equation

The Hamilton-Jacobi-Bellman equation is a PDE that plays a central role in optimal control theory, where allows to find the *value function* for a dynamical system with an associated cost function as well as it can be used to solve classical *brachistocrone* problems. The equation can be extended to stochastic systems and in general to a broad spectrum of problems. Solving the HJB equation is not trivial and is usually performed backward in time, however in the general case this PDE does not have a smooth solution, [3] therefore generalized solutions are used and several approximate methods have been proposed, also employing multi-layer perceptrons (approximated dynamic programming). [8] Although, until recently, high dimensional problems involving this PDE basically remained intractable and the famous term "curse of dimensionality" was firstly used by Richard Bellman right in the context of dynamical programming [21]. As a consequence of this, it is particularly interesting showing the ability of the PINN approach with quadrature methods in solving high dimensional HJB equations.

Let's consider a modified version of the example presented by [21] at page 6 that consists in a linear-quadratic-Gaussian control problem with $t \in [0, t]$, $X_0 = x$ and the cost functional $J(\{m_t\}_{t \in [0,T]}) = \mathbb{E}[\int_0^T ||m_t||^2 dt + g(X_t)]$ :

$$dX_t = 2\sqrt{\lambda} m_t dt + \sqrt{2} dW_t \tag{84}$$

In this formula $X_t$ represents the state process, $m_t$ the control process, $\lambda > 0$ is a constant representing the "strength" of the control and $W_t$ is a standard Brownian motion. The HJB equation for this process is given by

$$\frac{\partial u}{\partial t} + \nabla^2 u - \lambda ||\nabla u||^2 = 0 \tag{85}$$

As said before, the HJB equation is usually solved backward in time, therefore in this example the PDE will be solved in 5 dimensions (4 spatial + 1 temporal) with the following terminal condition:

$$u(t, x_1, x_2, x_3, x_4) = ln((1 + x_1^2 + x_2^2 + x_3^2 + x_4^2)/2) \tag{86}$$

For more details about the physical meaning of this condition please refers to [21].

Again, our model consists of a feed-forward NN (as in the previous case) with one input layer of 5 neurons, one hidden layer of 20 neurons and an output layer of dimension one. The input and the hidden layer have a sigmoid activation function, the output layer has a linear output.

The results are similar to those obtained in the previous example, as Fig. 18 and Fig. 20 show. GridTraining takes a time of the order of 5 and a half hours to perform one thousand iterations, making it impractical for actual use. On the other hand, the Quadrature based algorithms and QuasiRandomTraining strategy perform exceedingly well on the time benchmark while achieving a good performance on the convergence of the loss. StochasticTraining, which usually guarantees satisfactory results in terms of training time, has a comparable behaviour to QuasiRandomTraining.

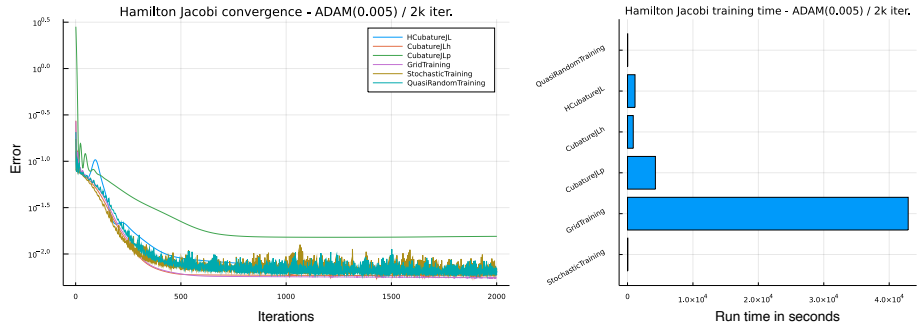

Figure 18: Convergence of the Hamilton Jacobi equation training using ADAM as minimizer.

| Strategy | Minimum Error Value |
| --- | --- |
| HCubatureJL | $5.92 \times 10^{-3}$ |
| CubatureJLh | $5.71 \times 10^{-3}$ |
| CubatureJLp | $1.51 \times 10^{-2}$ |
| GridTraining | $5.51 \times 10^{-3}$ |
| StochasticTraining | $5.90 \times 10^{-3}$ |
| QuasiRandomTraining | $5.85 \times 10^{-3}$ |

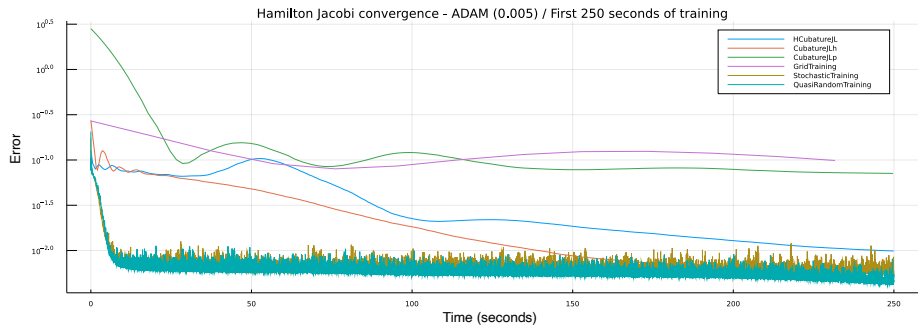

Figure 19: Training convergence with time on the horizontal axis (first 250 seconds of training).

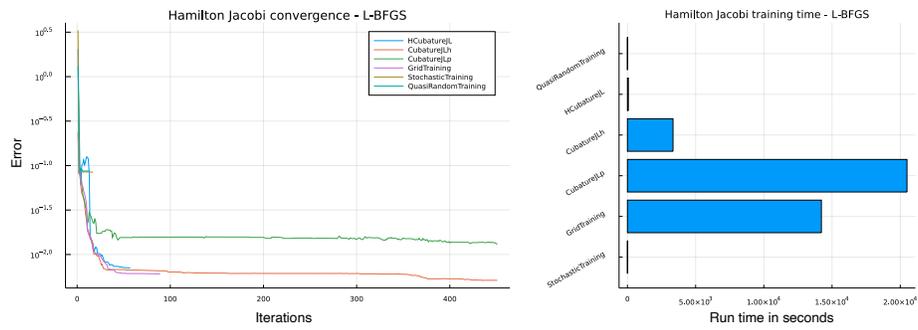

Figure 20: Convergence of the Hamilton Jacobi equation training using L-BFGS as minimizer.

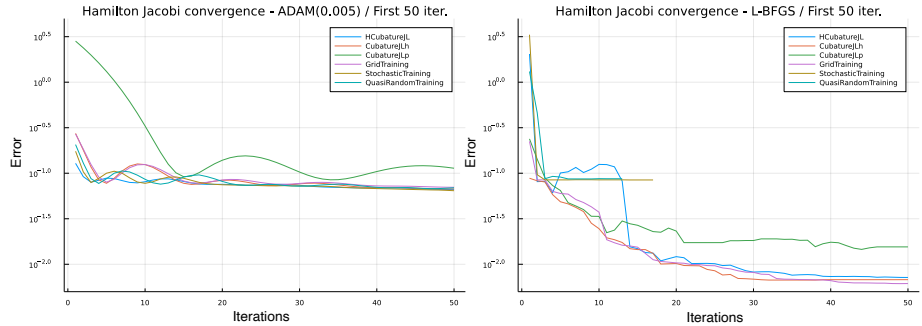

Figure 21: Convergence speed comparison between ADAM and L-BFGS.

## 4.7 Comparing Different Optimizers

This section is a case study comparing the results of different optimizers on the models described by the above examples. The training strategy used in all cases is *CubatureJLh* with *reltol= 1e-6*, *abstol= 1e-3* and *maxiters= 1e3*. The cases tested for were:

1. ADAM(0.001) for 200 maximum iterations.
2. ADAM(0.005) for 200 maximum iterations.
3. ADAM(0.05) for 200 maximum iterations.
4. RMSProp(0.001) for 200 maximum iterations.
5. RMSProp(0.005) for 200 maximum iterations.
6. RMSProp(0.05) for 200 maximum iterations.
7. BFGS() for 200 maximum iterations.
8. LBFGS() for 200 maximum iterations.
9. ADAM(0.001) for 50 iterations followed by BFGS() for 150 iterations.

### 4.7.1 1-D Diffusion

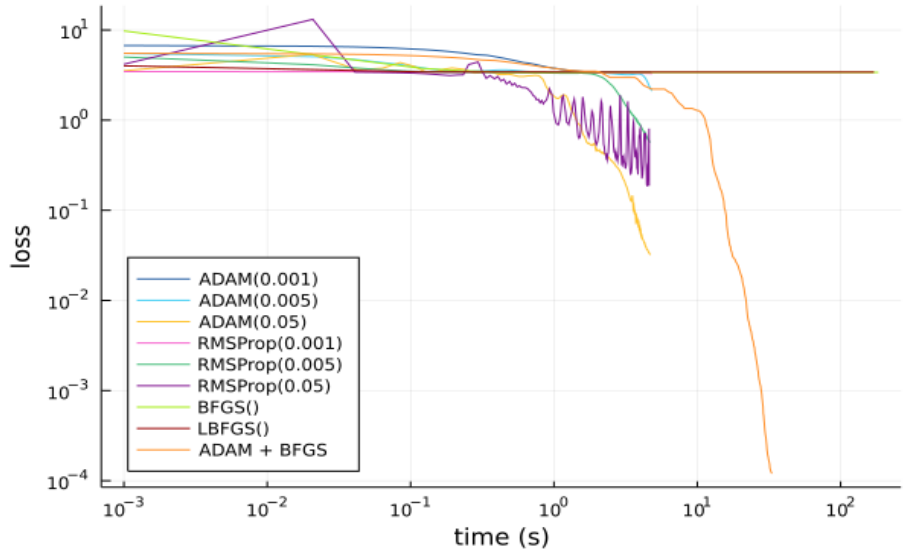

Figure 22: 1-D Diffusion: Loss Over Time

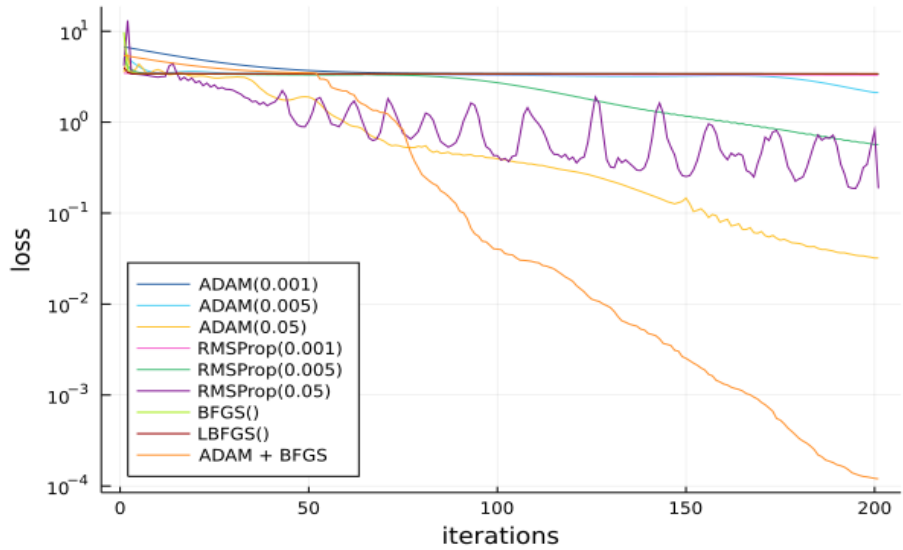

Figure 23: 1-D Diffusion: Loss Over Iterations

### 4.7.2 Simple Poisson Example

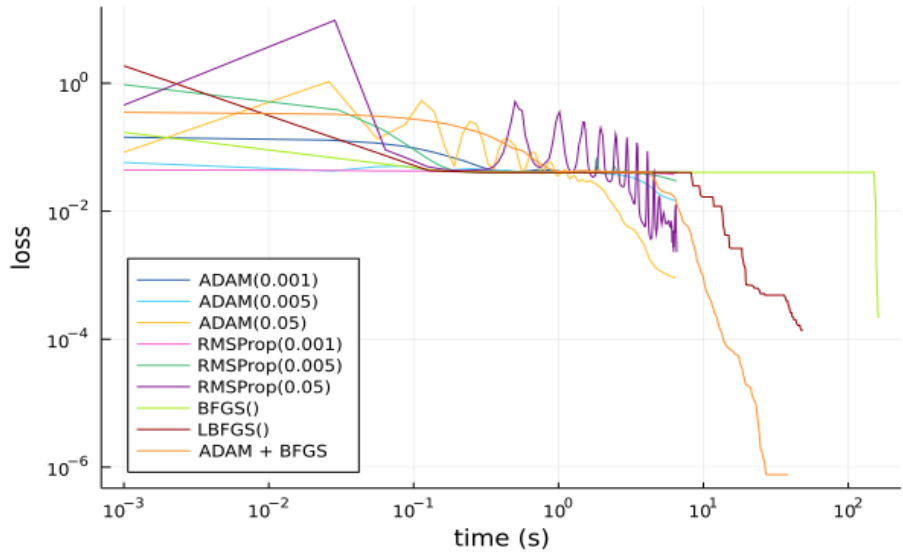

Figure 24: Simple Poisson Example: Loss Over Time

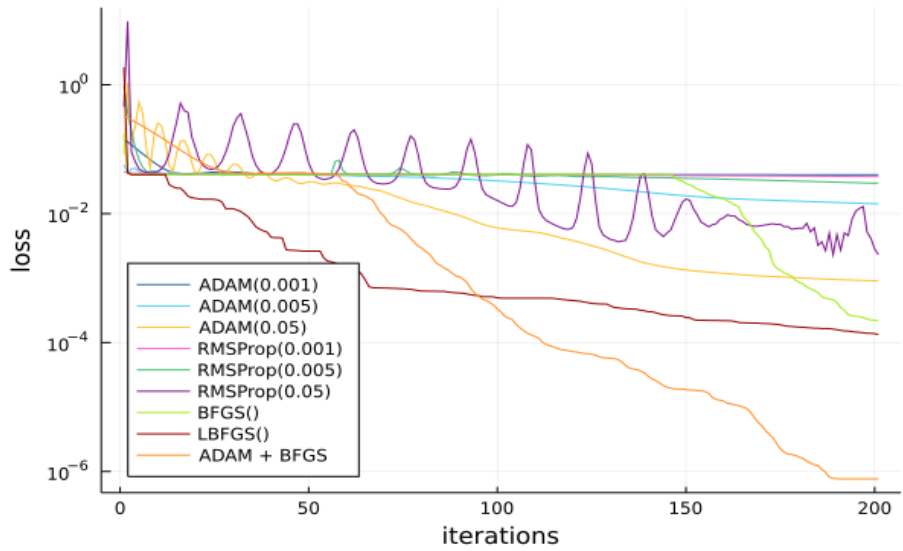

Figure 25: Simple Poisson Example: Loss Over Iterations

### 4.7.3 Poisson-Nernst-Planck

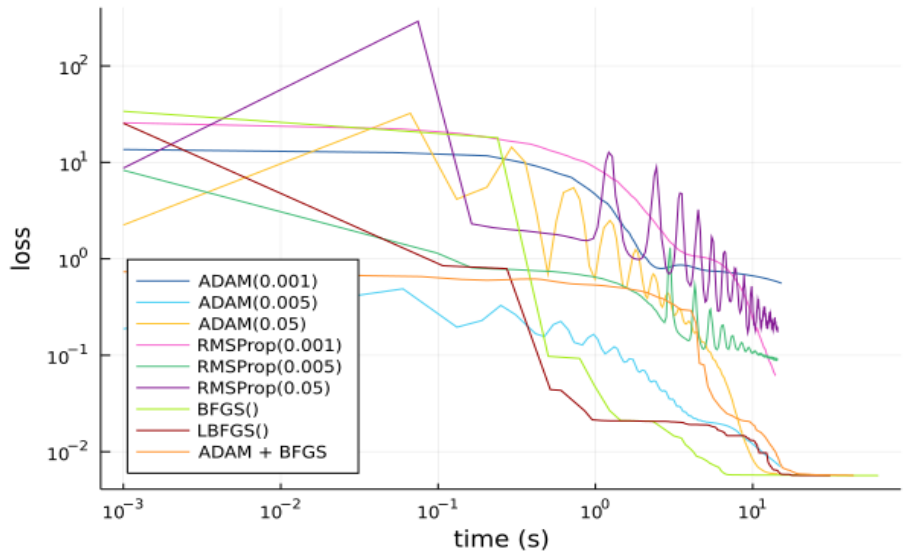

Figure 26: Poisson-Nernst-Planck: Loss Over Time

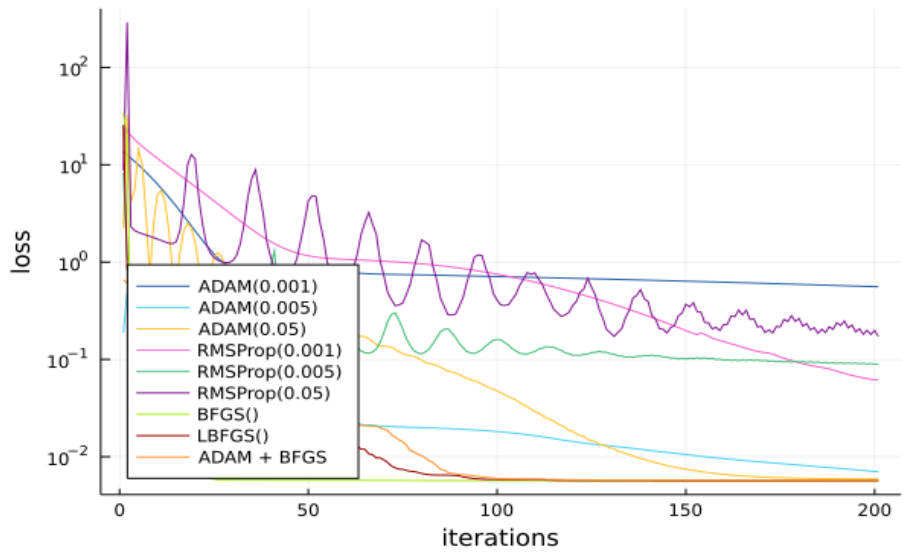

Figure 27: Poisson-Nernst-Planck: Loss Over Iterations

### 4.7.4 Burger's Equation

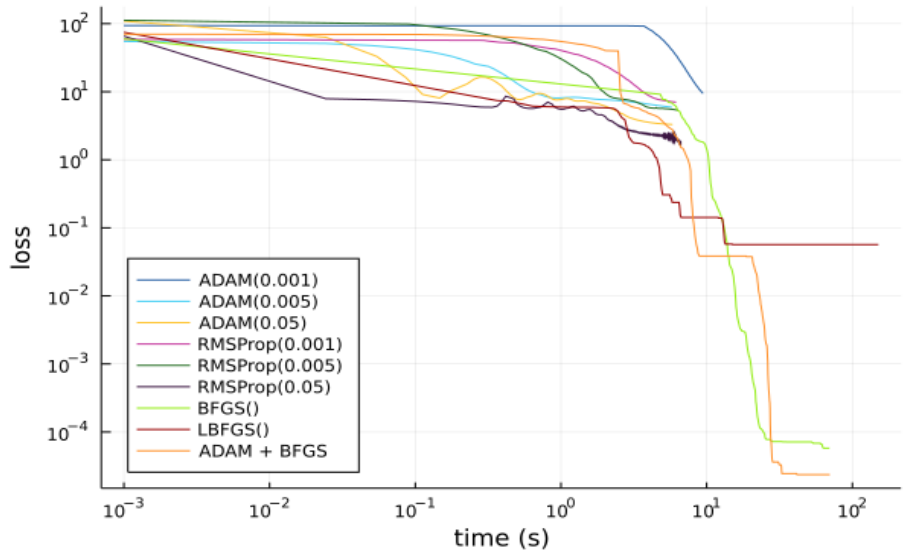

Figure 28: Burger's Equation: Loss Over Time

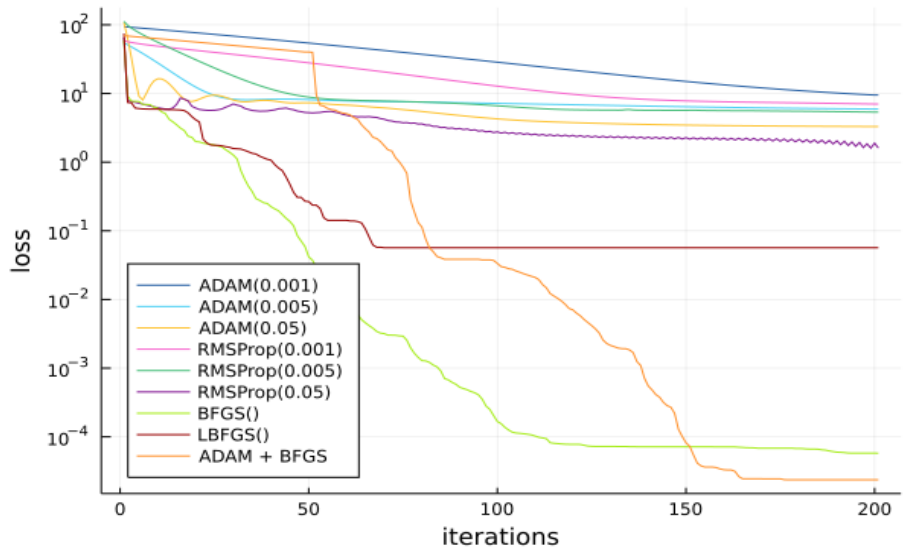

Figure 29: Burger's Equation: Loss Over Iterations

### 4.7.5 Allen Cahn Equation

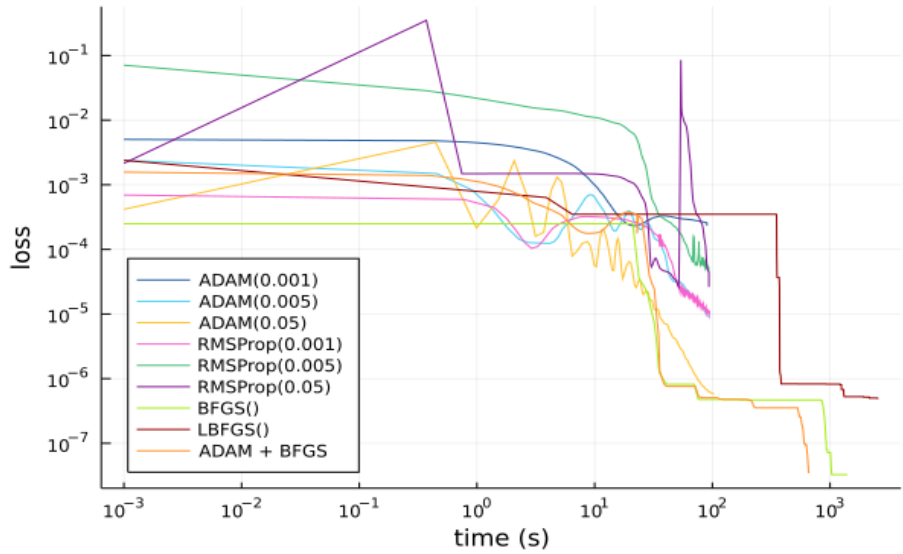

Figure 30: Allen Cahn Equation: Loss Over Time

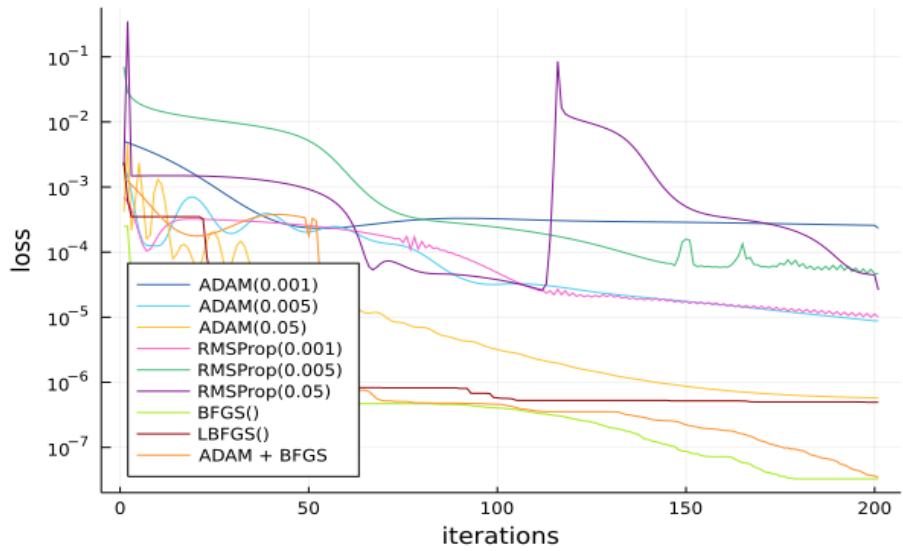

Figure 31: Allen Cahn Equation: Loss Over Iterations

### 4.7.6 Hamilton-Jacobi-Bellman Equation

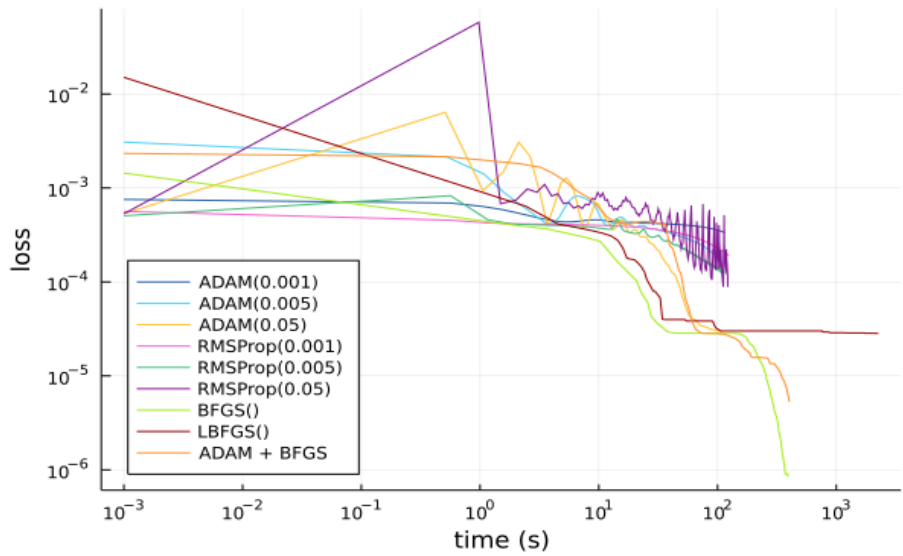

Figure 32: Hamilton-Jacobi-Bellman Equation: Loss Over Time

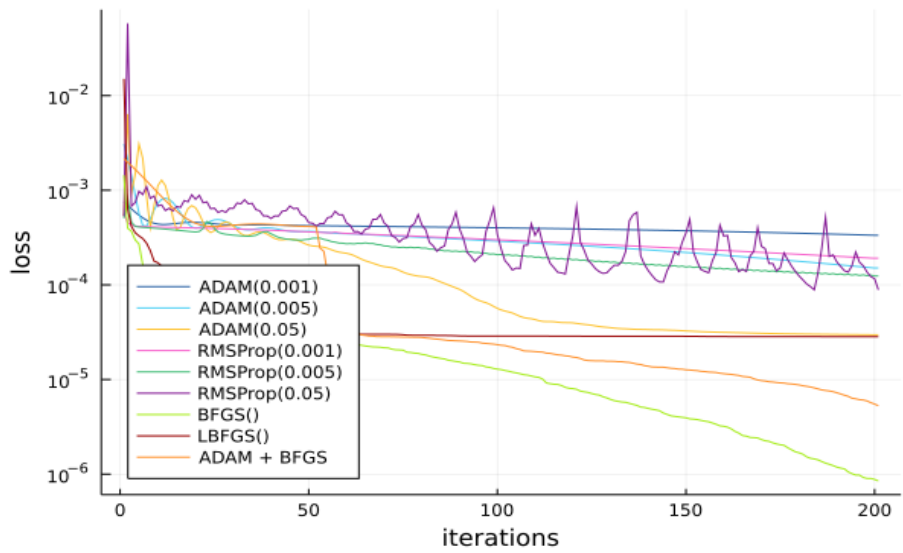

Figure 33: Hamilton-Jacobi-Bellman Equation: Loss Over Iterations

The following table depicts the minimum loss values for each case:

| Opt. | Diffusion. | Poisson. | PNP. | Burger's. | Allen. | Hamilton. |
|---|---|---|---|---|---|---|
| ADAM(0.001) | 3.387 | 0.040 | 0.559 | 9.493 | 0.0002 | 0.0003 |
| ADAM(0.005) | 2.120 | 0.014 | 0.007 | 5.935 | $8.8e-6$ | 0.0001 |
| ADAM(0.05) | 0.032 | 0.0009 | 0.005 | 3.299 | $5.7e-7$ | $2.9e-5$ |
| RMSProp(0.001) | 3.297 | 0.037 | 0.061 | 7.022 | $9.9e-6$ | 0.0002 |
| RMSProp(0.005) | 0.567 | 0.029 | 0.089 | 5.348 | $4.6e-5$ | 0.0001 |
| RMSProp(0.05) | 0.187 | 0.002 | 0.172 | 1.602 | $2.6e-5$ | $8.8e-5$ |
| BFGS() | 3.390 | 0.0002 | 0.005 | $5.7e-5$ | $3.2e-8$ | $8.4e-7$ |
| LBFGS() | 3.412 | 0.0001 | 0.005 | 0.056 | $4.9e-7$ | $2.8e-5$ |
| ADAM+BFGS | 0.0001 | $7.6e-7$ | 0.005 | $2.3e-5$ | $3.4e-8$ | $5.2e-6$ |

These results reinforce the general observation that ADAM and RMSProp converge slower than BFGS, but BFGS often gets stuck at a local optima. A few iterations of ADAM narrow down the loss enough so that the rest of the iterations of BFGS would converge rapidly towards a global optima. Therefore, the hybrid case proposes the best combination of optimizers and gives optimal results for a given number of iterations. However, plain BFGS yields better results in cases where it does not get stuck at a local optima.

# 5 Summary of Numerical Experiment Results

## 5.1 Summary of Loss Discretization Strategies

One result which was unable to be conveyed by the numerical experiments is that the adaptive quadrature techniques typically showed the most robustness in solving the problem, easily adapting to new problems in a straightforward manner while in some cases hyperparameter tuning was required to make stochastic or quasi-random sampling work. However, in many cases the extra robustness of the quadrature-based techniques came at a cost. When plotting error vs time the quadrature techniques can many times be slower. Looking at the points it was adaptively focusing on, this seems to be due to spending a lot of computational effort to accurately calculate the loss even when it's unnecessary. For example, when the initial condition is incorrect, it would not make sense to very accurately calculate the loss at later time periods, but the quadrature method will spend even more time in these later regions since the instabilities will cause them to be more incorrect and thus adaptively gather more evaluation points. Methods which adaptively change the tolerance of the quadrature as the optimization iterates could be an effective strategy to explore.

From these results we see that, as expected, the `GridStrategy` generally does not perform well in terms of robustness or speed as the dimensionality of the problem increases. One of the best performers is the quasi-random sampling technique.

## 5.2 Summary of Optimizer Choices

Tests with (L-)BFGS show that the method is more prone to arriving in local minima than optimizers like ADAM, but tends to hit lower loss values in significantly fewer iterations if it is not locked in such a minima. Many of the plots show a classic sign of a high learning rate on ADAM, the periodic spikes to higher losses, which users need to be aware of. While the spikes can be an issue, we noticed in the testing that allowing for a slightly spiky learning rate can still significantly improving the optimization speed while not reducing the long term robustness of the optimization. Though users are advised to use caution when seeing this behavior.

# 6 The Doyle-Fuller-Newman Model

In this section, we present simulations of the Doyle-Fuller-Newman (DFN) model [13], a model which is commonly used to simulate batteries at the cell scale, incorporating mass transport within particles and electrolyte, reaction kinetics at the particle surface, and distribution of electric potentials. Typically, the goal of this model is to output the voltage given the current (or vice versa) as functions of time, but spatially distributed quantities such as particle concentrations, electrolyte concentrations, and distributions of currents and potentials may also be of interest (though harder or impossible to measure experimentally).

The DFN presents a number of implementation challenges making it a good test for the NeuralPDE algorithms:

- different variables defined in different subdomains either at the macroscale (negative electrode, separator, positive electrode) or the microscale (negative particles, positive particles)

- coupling between microscale and macroscale via the surface value and boundary condition of the microscale variable (particle concentration)

- highly nonlinear source term coupling electrolyte concentration and solid- and liquid-phase electric potentials

Since the DFN model is a homogenized model that couples together the microscale dimension of the particles and the macroscale dimension of the cell, it is sometimes also referred to as the pseudo-two-dimensional (P2D) model.

Due to the complexity of the model, several ways of simplifying the model have been explored. The most popular are the Single Particle Model (SPM) and Single Particle Model with electrolyte (SPMe), which are valid in the limit of fast diffusion in the electrolyte [34] (Section 6.2). These models remove the coupling between microscale and macroscale and are therefore much easier to implement and solve. Another – less physically-relevant but numerically educational – way to simplify the model is to ignore the microscale and only solve for macroscale variables such as electrolyte concentration, electrolyte potential, and solid-phase potential (Section 6.3).

The results of the P2D model will depend on the parameter set being used, as the model can be used to simulate any type of lithium-ion battery. In these examples we use the parameter set from Marquis et al. [34].

## 6.1 Full pseudo-2D model

Before we state the governing equations we comment on our notation. To distinguish variables in the electrolyte from those in the solid phase of the electrode, we use a subscript e for electrolyte variables and a subscript s for solid phase variables. To indicate the region within which each variable is defined, we include an additional subscript k, which takes one of the following values: n (negative electrode), p (positive electrode), or s (separator). For example, the notation $\phi_{s,n}$ refers to the potential in the solid phase of the negative electrode.

**Mass conservation in the active material.** As in [13], we treat the active material on the microscale as spherical particles of uniform radius, in which spherically symmetric diffusion of lithium is described by Fick's law. Mass conservation (for k ∈ {n, p}) gives

$$\frac{\partial c_{s,k}}{\partial t} = -\frac{1}{r^2}\frac{\partial}{\partial r}\left(r^2 N_{s,k}\right), \quad N_{s,k} = -D_{s,k}(c_{s,k}, T_k)\frac{\partial c_{s,k}}{\partial r}. \tag{87a}$$

where $c_{s,k}$ is the concentration of lithium in the active material, $N_{s,k}$ is the flux of lithium ions in the active material, $D_{s,k}(c_{s,k}, T_k)$ is the diffusivity of lithium in the active material, $T_k$ is the (macroscopic) temperature, $r$ is the radial spatial coordinate, and $t$ is time. We assume that the particle is entirely surrounded by electrolyte and that lithium transfer with the electrolyte occurs uniformly across each particle's outer surface, giving

$$N_{s,k}\big|_{r=0} = 0, \quad N_{s,k}\big|_{r=R_k} = \frac{j_k}{F}, \tag{87b}$$

where $j_k$ is the interfacial current density, $R_k$ is the particle radius, and $F$ is Faraday's constant. Further, we assume that the concentration within the particles in each electrode is initially uniform in space

$$c_{s,k}\big|_{t=0} = c_{s,k,0}, \tag{87c}$$

where $c_{s,k,0}$ is a constant.

**Electrochemical reactions.** The electrochemical reactions are modelled using symmetric Butler–Volmer kinetics. The interfacial current density $j_k$ (for k ∈ {n, p}) is then given by

$$j_k = 2j_{0,k}(c_{s,k}, c_{e,k}, T)\sinh\left(\frac{F\eta_k}{2R_g T}\right), \tag{88a}$$

$$\eta_k = \phi_{s,k} - \phi_{e,k} - U_k(c_{s,k}, T)\big|_{r_k=R_k}, \tag{88b}$$

where $c_{e,k}$ is the concentration of lithium-ions in the electrolyte, $j_{0,k}(c_{s,k}, c_{e,k}, T)$ is the exchange current density, $\eta_k$ is the surface overpotential, $U_k$ is the open circuit potential, and $R_g$ is the universal gas constant. In the separator region there is no active material so the interfacial current density is zero, i.e. $j_s = 0$.

**Charge conservation in the electrodes.** The current in the electrodes is described by Ohm's law. Charge conservation (for $k \in \{n, p\}$) then gives

$$i_{s,k} = -\frac{\partial \phi_{s,k}}{\partial x}, \quad \frac{\partial i_{s,k}}{\partial x} = -a_k j_k, \tag{89a}$$

where $x$ is the spatial coordinate in the macroscopic dimension, $i_{s,k}$ is the current density, $\phi_{s,k}$ is the electric potential, and $a_k$ is the surface area per unit volume of the electrode. The term $-a_k j_k$ accounts for transfer of current between the electrode and electrolyte, which occurs via electrochemical reactions. We choose set set the reference potential to be $0\,\text{V}$ at $x = 0$

$$\phi_{s,n}\big|_{x=0} = 0. \tag{89b}$$

The separator is taken to be electrically insulating so that no charge is transferred from the electrodes to the separator

$$i_{s,n}\big|_{x=L_n} = 0, \quad i_{s,p}\big|_{x=L_n+L_s} = 0. \tag{89c}$$

The remaining boundary condition is provided by either specifying the current

$$i_{s,p}\big|_{x=L_x} = \frac{I_{\text{app}}}{A}, \tag{89d}$$

where $I_{\text{app}}(t)$ is the applied current, and $A$ is the cross-sectional area of the electrode, or the potential

$$\phi_{s,p}\big|_{x=L_x} = V, \tag{89e}$$

where $V(t)$ is the applied voltage. Alternatively, one could prescribe the power or some combination of current, voltage or power control (e.g. by specifying an experimental protocol such as CCCV).

**Charge conservation in the electrolyte.** In the electrolyte current is driven by concentration gradients and the effects of interacting species must be accounted for, resulting in a modified Ohm's Law (for k ∈ {n, s, p})

$$i_{e,k} = \kappa^{\text{eff}}(c_{e,k}, T) \left( -\frac{\partial \phi_{e,k}}{\partial x} + 2(1 - t^+) \frac{RT}{F} \frac{\partial}{\partial x} \log(c_{e,k}) \right), \tag{90a}$$

where $i_{e,k}$ is the current in the electrolyte, $\kappa^{\text{eff}}(c_{e,k}, T_k)$ is the electrolyte conductivity, $\phi_{e,k}$ is the electrical potential in the electrolyte, and $t^+$ is the transference number. In the electrodes charge is transferred between the electrolyte and the electrode solid material, whereas n the separator no charge transfer occurs between the electrolyte and the separator material. This gives (for k ∈ {n, s, p})

$$\frac{\partial i_{e,k}}{\partial x} = a_k j_k, \tag{90b}$$

where $j_s = 0$. No charge is transferred directly from the electrolyte into the current collectors, so that

$$i_{e,n}\big|_{x=0} = 0, \quad i_{e,p}\big|_{x=L_x} = 0. \tag{90c}$$

Further, both the potential and current must be continuous at the electrode/separator interfaces.

**Mass conservation in the electrolyte.** The concentration of lithium ions in the electrolyte is determined by solving an effective diffusion equation with an additional source term describing lithium transfer to the active material (for k ∈ {n, s, p})

$$\epsilon_k \frac{\partial c_{e,k}}{\partial t} = -\frac{\partial N_{e,k}}{\partial x} + \frac{1-t^+}{F} a_k j_k, \tag{91a}$$

$$N_{e,k} = -D^{\text{eff}}(c_{e,k}, T_k) \frac{\partial c_{e,k}}{\partial x}, \tag{91b}$$

where $\epsilon_k$ is the electrolyte volume fraction, $N_{e,k}$ is the lithium-ion flux in the electrolyte, and $D^{\text{eff}}(c_{e,k}, T_k)$ is the diffusivity of the electrolyte. There is no flux of lithium ions from the electrolyte into the current collectors

$$N_{e,n}\big|_{x=0} = 0, \quad N_{e,p}\big|_{x=L_x} = 0. \tag{91c}$$

Further, we require continuity of concentration and flux at the electrode/separator interfaces. Finally, we assume that the concentration of lithium ions in the electrolyte is initially uniform in space

$$c_{e,k}\big|_{t=0} = c_{e,0}, \tag{91d}$$

where $c_{e,0}$ is constant.

### 6.2 Example 8: SPM

In this section we discuss experiments with using PINNs to learn the Single Particle Model (SPM), which is a much simplified version of the DFN model. This model consists of three PDE losses and seven boundary condition losses, which makes it more complicated than most of the models considered thus far:

$$\frac{\partial Q(t)}{\partial t} = 4.27249308415467 \tag{92}$$

$$\frac{\partial c_{s_n xav}(t, r_n)}{\partial t} = \frac{8.813457647415216}{r_n^2} \frac{\partial(r_n^2(\partial c_{s_n xav}(t, r_n)))}{\partial r_n^2} \tag{93}$$

$$\frac{\partial c_{s_p xav}(t, r_p)}{\partial t} = \frac{22.598609352346717}{r_p^2} \frac{\partial(r_p^2(\partial c_{s_p xav}(t, r_p)))}{\partial r_p^2} \tag{94}$$

$$Q(0) = 0, \tag{95}$$

$$c_{s_n xav}(0, r_n) = 0.8, \tag{96}$$

$$c_{s_p xav}(0, r_p) = 0.6, \tag{97}$$

$$\frac{\partial c_{s_n xav}(t, 0)}{\partial r_n} = 0, \tag{98}$$

$$\frac{\partial c_{s_n xav}(t, 1)}{\partial r_n} = -0.14182855923368468, \tag{99}$$

$$\frac{\partial c_{s_p xav}(t, 0)}{\partial r_p} = 0, \tag{100}$$

$$\frac{\partial c_{s_p xav}(t, 1)}{\partial r_p} = 0.03237700710041634, \tag{101}$$

We train a NeuralPDE PINN with a nonadaptive loss, where each of the boundary conditions and pde losses are given equal, constant weight (see 2.10), and display the learned functions for the concentrations with respect to $r_n$ and $r_p$ below. For the experiments in this section and the next section, we used the same training settings except for the adaptive reweighting algorithm, to more directly compare the effect of the adaptive reweighting schemes. We used neural networks with hidden layers of 50 dimensions, and 2 hidden layers, with GELU nonlinearities. Integration was performed using QuadratureTraining with the HCubatureJL cubature method, an abstol of $1e-5$, reltol of 1, and maxiters of 1000. Optimization was performed via 50,000 iterations of ADAM with a learning rate of $3e-4$. We compare the learned functions against the Python Battery Mathematical Modelling (PyBaMM) library to act as roughly ground truth finite different simulation of this model and show the absolute error.

We do not display or compare the error on the $Q(t)$ learned function, as it is simply a linear function, and all models were able to essentially perfectly learn this behavior.

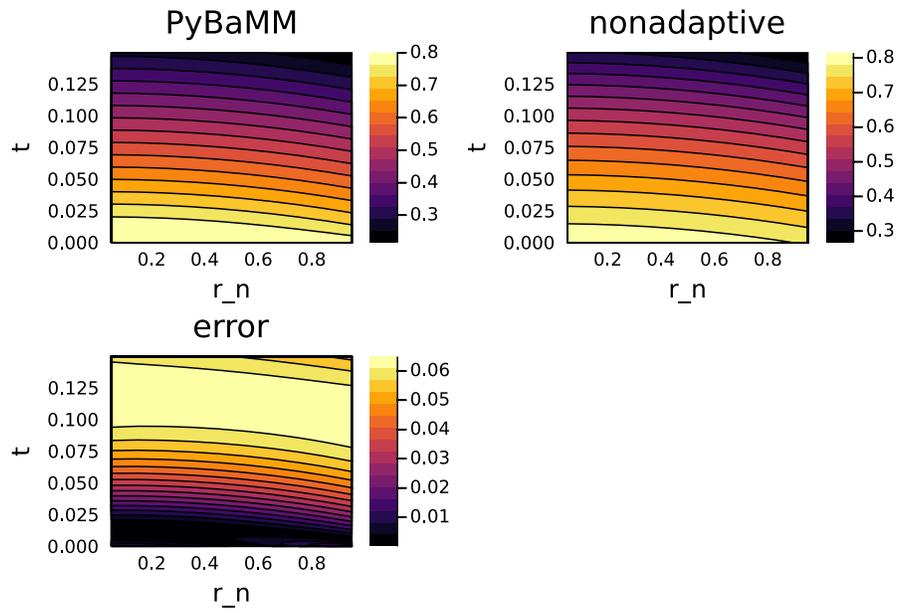

Figure 34: Comparison of the nonadaptive PINN's prediction with PyBaMM solution for $c_{s_n xav}$, with 8.74% relative L2 error.

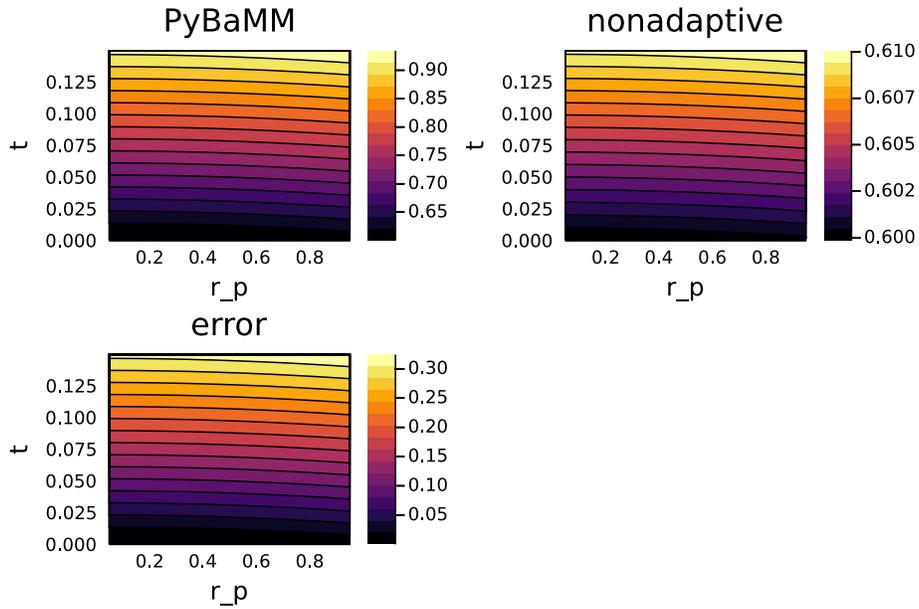

Figure 35: Comparison of the nonadaptive PINN's prediction with PyBaMM solution for $c_{s_p xav}$, with 23.67% relative L2 error.

We see that the nonadaptive PINN 34, 35 is able to learn certain features, for instance the curvature of the solution, but is unable to match all of the features of the solution function simultaneously. The error for $c_{s_n xav}$ is not extremely large, at 8.74% relative L2 error, but there is a large error for the learned function for $c_{s_p xav}$, with 23.67% relative L2 error.

These errors can be largely mitigated by using one of the two adaptive reweighted loss functions, the Loss Gradients adaptive reweighting or the MiniMax adaptive reweighting.

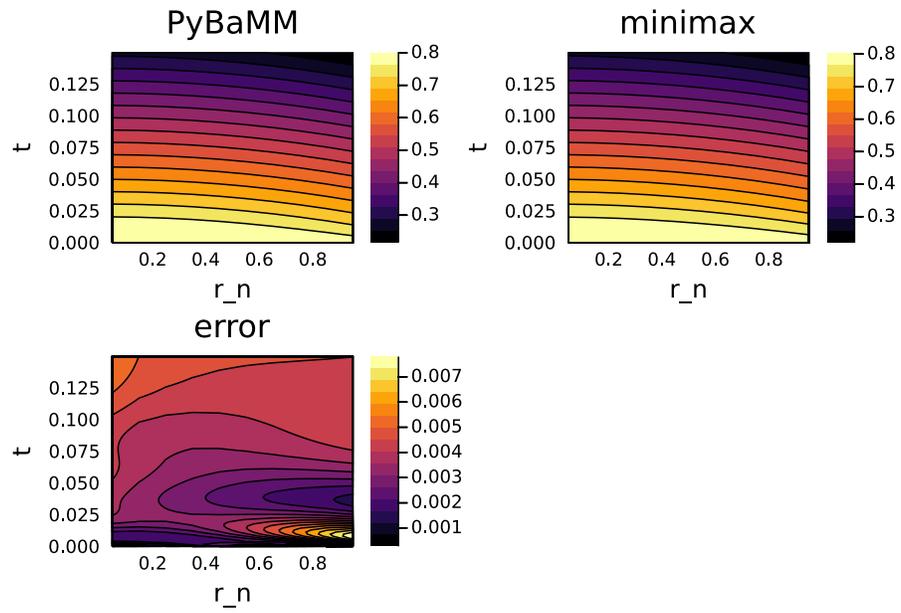

Figure 36: Comparison of the minimax PINN's prediction with PyBaMM solution for $c_{s_n x a v}$, with 0.68% relative L2 error.

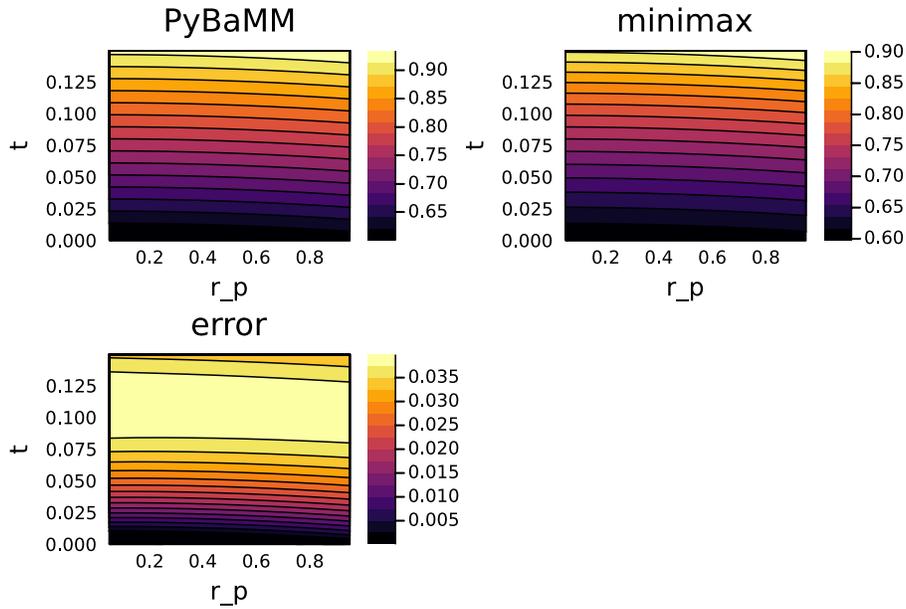

Figure 37: Comparison of the minimax PINN's prediction with PyBaMM solution for $c_{s_p xav}$, with 4.06% relative L2 error.

When MiniMax adaptive reweighting is used, 36, 37 we see that the learned function matches the solution extremely closely, with only 0.68% relative L2 error for $c_{s_n xav}$ and 4.06% relative L2 error for $c_{s_p xav}$. There is a persistent drift in the $c_{s_p xav}$ learned function, and it is unclear why this drift exists and stays present across different training runs.

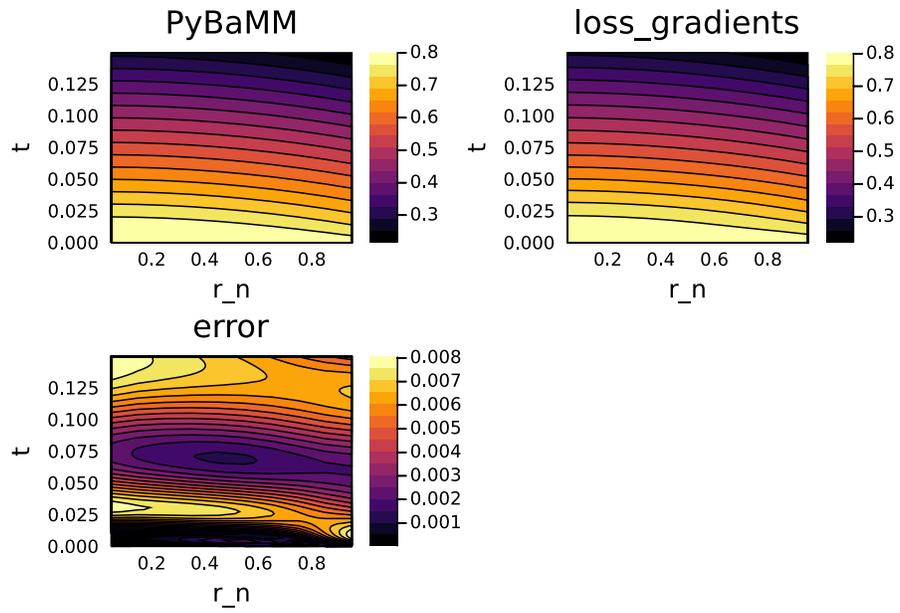

Figure 38: Comparison of the loss gradients PINN's prediction with PyBaMM solution for $c_{s_n xav}$, with 0.88% relative L2 error.

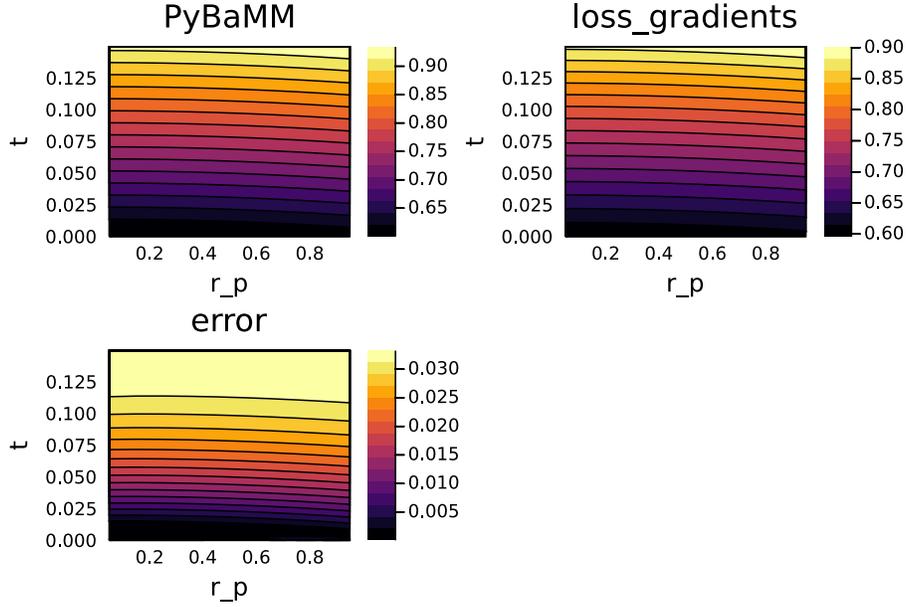

Figure 39: Comparison of the loss gradients PINN's prediction with PyBaMM solution for $c_{s_p xav}$, with 3.1% relative L2 error.

With the Loss Gradients adaptive reweighting, we see a very similar performance to the MiniMax reweighting 38, 39. The $c_{s_n xav}$ learned function has only only 0.88% relative L2 error and 3.1% relative L2 error for $c_{s_p xav}$. We see a very similar drift compared to the simulation for $c_{s_p xav}$ using Loss Gradients adaptive reweighting as we did when using MiniMax reweighting, suggesting that this might have something to do with the problem formulation.

## 6.3 Example 9: Reduced P2D

Let's take a look at the reduced Pseudo-two-Dimensional(P2D) model - systems of two equations:

$$\frac{\partial c_e(t,x)}{\partial t} = \frac{\partial^2 c_e(t,x)}{\partial x^2} + f(x) \tag{102}$$

$$0 = \frac{\partial^2 c_e(t,x)}{\partial x^2} - \frac{\partial^2 \phi_e(t,x)}{\partial x^2} - f(x) \tag{103}$$

where $c_e$ - electrolyte concentration, $\phi_e$ - electrolyte potential and f - electrolyte concentration equation:

$$f(x) = \begin{cases} n, & \text{if } x \in L_n \\ s, & \text{if } x \in L_s \\ p, & \text{if } x \in L_p \end{cases} \tag{104}$$

- n, s, p - concatenation in the electrolyte domain,
- $L_n, L_s, L_p$ - negative, separator, positive domains,

with initial and boundary conditions:

$$c_e(0, x) = 1, \tag{105}$$
$$\phi_e(0, x) = 0.0, \tag{106}$$
$$\frac{\partial c_e(t, 0.0)}{\partial x} = 0, \tag{107}$$
$$\frac{\partial c_e(t, 1.0)}{\partial x} = 0, \tag{108}$$
$$\phi_e(t, 0.0) = 0, \tag{109}$$
$$\frac{\partial \phi_e(t, 1.0)}{\partial x} = 0.0 \tag{110}$$
$$\tag{111}$$

Figures 40, 41 below demonstrate the results of predicting the solution using NeuralPDE PINNs and comparing them with the results of PyBaMM (Python Battery Mathematical Modelling).

These are a set of highly coupled and nonlinear equations and so are difficult to learn for any of the methods discussed so far. The nonadaptive method is able to learn a function for $c_e$ with good accuracy 40, but fails completely at learning the $\phi_e$ 41 function, which is tightly coupled throughout learning to the $c_e$ function.

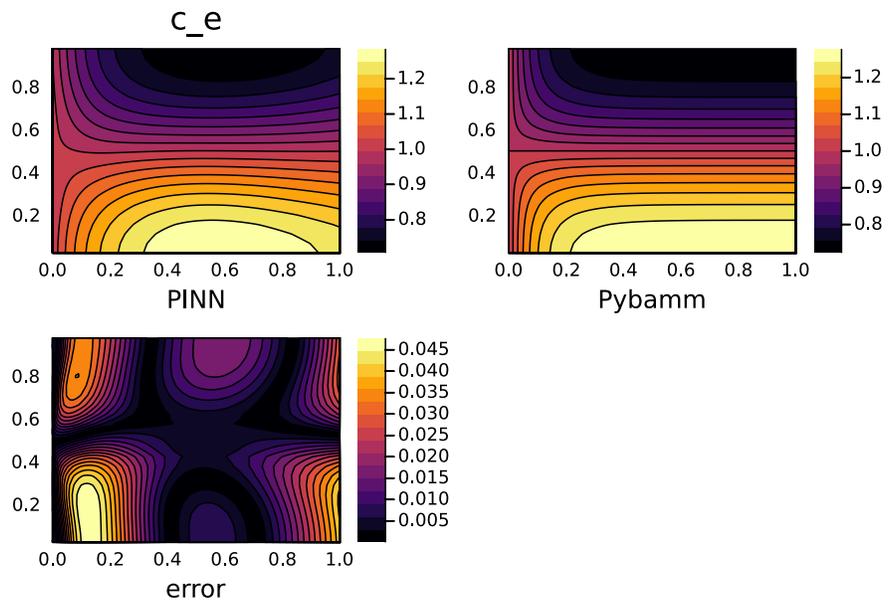

Figure 40: Comparison of the nonadaptive PINN's prediction with PyBaMM solution for $c_e$.

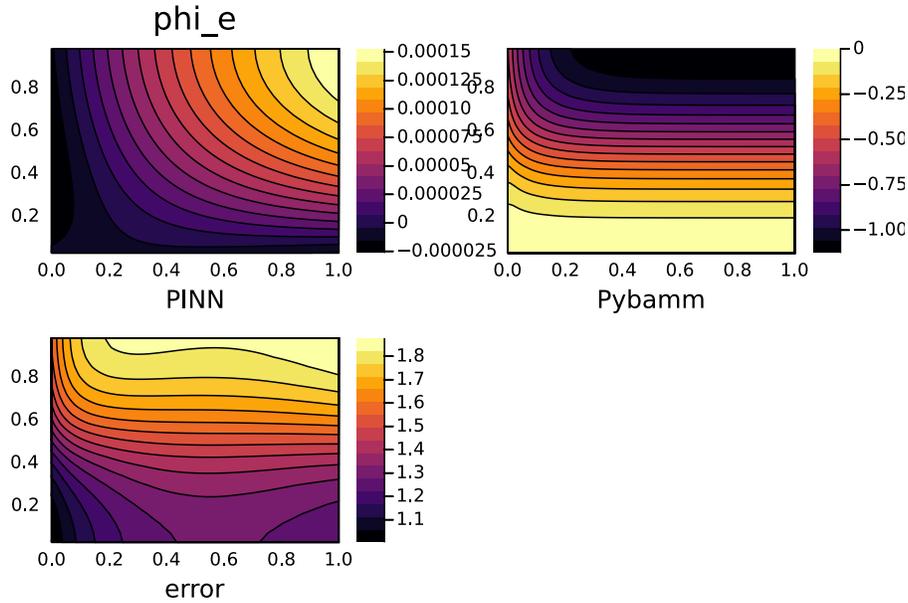

Figure 41: Comparison of the nonadaptive PINN's prediction with PyBaMM solution for $\phi_e$.

The Loss Gradients method, interestingly, was able to learn a somewhat accurate $\phi_e$ 43, while learning an incredibly inaccurate $c_e$ 42 function. It is unclear why this occurred, but during training this method very often has one or more of the adaptive weight quickly converge to the maximum allowed weight as that loss gradient approaches a local minima or maxima, because then the loss gradient for that term approaches 0. Note that we do implement a maximum adaptive weight by performing the exponential moving average of weights by $\frac{1}{mean(\nabla C_i(w))+1e-7}$, where the $1e-7$ term becomes limiting as the mean of the gradient approaches 0. This is necessary to prevent asymptotic instability, but ends up becoming another user choice.

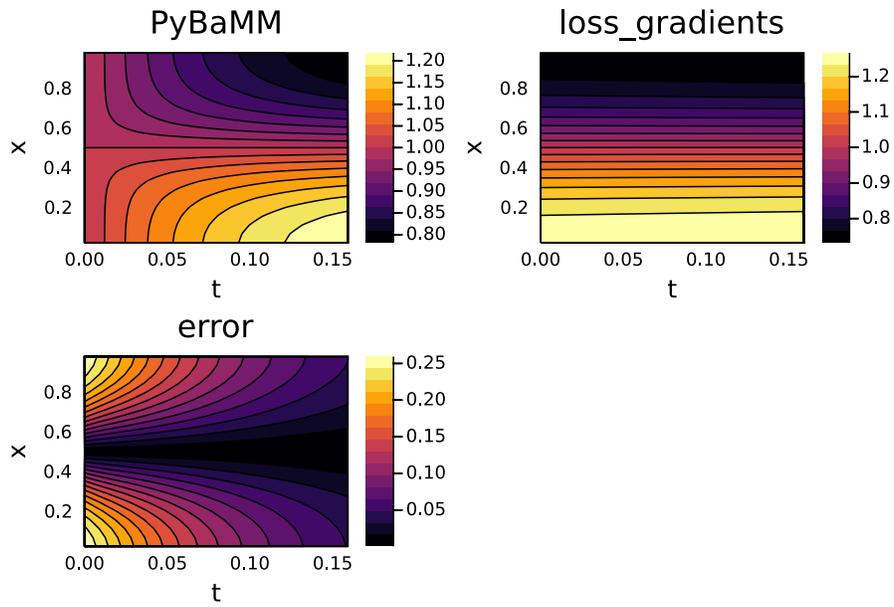

Figure 42: Comparison of the loss gradients PINN's prediction with PyBaMM solution for $c_e$, with 8.84% relative L2 error.

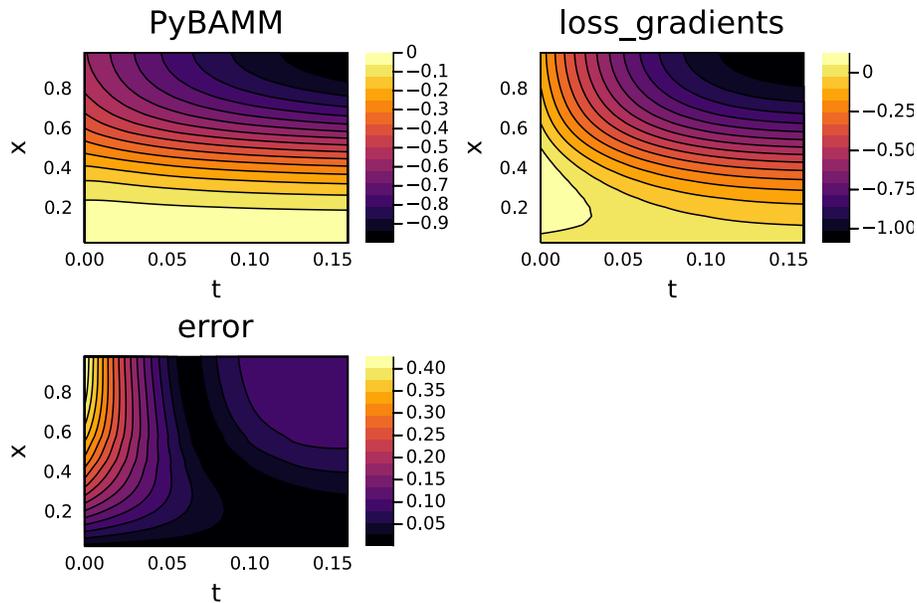

Figure 43: Comparison of the loss gradients PINN's prediction with PyBaMM solution for $\phi_e$, with 25.06% relative L2 error.

The MiniMax adaptive reweighting scheme fared much better and was able to learn an incredibly accurate $c_e$ 44 function, with only 0.12% relative L2 error. The learned function for $\phi_e$ 45 had very interesting behavior, where the error was almost entirely localized to one boundary, indicating that the second boundary condition, for $\phi_e$ at $t = 0$, was unsatisfied. This had a rippling effect in the error, placing it at 13.36% relative L2 error, despite being very close for much of the domain. It is unclear why this boundary condition in particular had such difficulty being satisfied when the rest of the model matched so well.

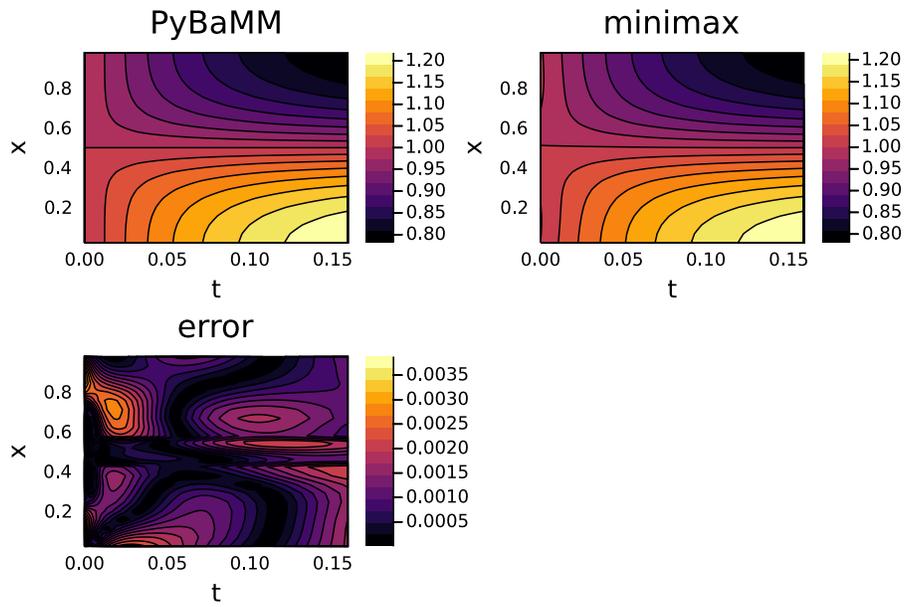

Figure 44: Comparison of the minimax PINN's prediction with PyBaMM solution for $c_e$, with 0.12% relative L2 error.

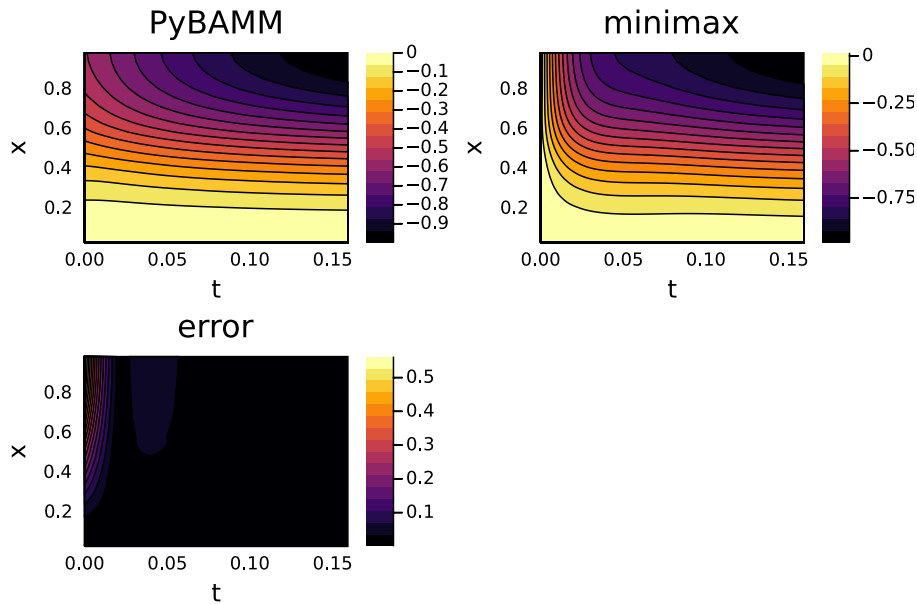

Figure 45: Comparison of the minimax PINN's prediction with PyBaMM solution for $\phi_e$, with 13.36% relative L2 error.

Future experiments will be aimed at honing these methods to reduce the error in learned PINN functions, and to make these methods more reliably and repeatably handle PDEs of the complexity found in battery models and other complicated PDE models. These improvements will be then used to expand these methods to learn models for the full P2D model, and then eventually to learn a single PINN model that has learned to simulate P2D models for a wide domain of parameters that we can specify in our PINN training process.

Another avenue of improvement is the incorporation of simulation data from the PDE such as what we compare against here via PyBaMM, into the training process. A neural network can be pretrained in a purely supervised fashion to predict the values of the PDE from the domain of the PDE by using the provided simulation data. This method is quite limited, as it cannot be relied on to extrapolate from known data, only interpolate, but it can be combined with the PINN approach as a pretraining step or as a regularizer to force the learned PINN model to agree with available simulation data and learn faster, and then use the physics-informed training process to extrapolate to the rest of the domain from that supervision simulation or real data.

We believe that with the combination of these two avenues of improvement, a neural P2D model that works for a wide range of input parameters and situations in an efficient, fast, feedforward manner, will be possible.

# 7 Reproducing the Results

All of the results for this paper can be reproduced using the tutorials of NeuralPDE.jl, along with the benchmark code in the following repository: https://github.com/ChrisRackauckas/PINN_Quadrature. We note that these benchmarks will be setup on the https://github.com/SciML/SciMLBenchmarks.jl platform in order to allow for automatic regeneration of the benchmarks and performance tracking over time.

# 8 Acknowledgements

This project was funded under the Ford-MIT Alliance. We thank Kaushik Balakrishnan and Devesh Upadhyay from Ford Research for discussions on PINNs and Batteries. The information, data, or work presented herein was funded in part by the Advanced Research Projects Agency-Energy (ARPA-E), U.S. Department of Energy, under Award Numbers DE-AR0001222, DE-AR0001211 and NSF grant OAC-1835443. And also in part by DARPA under agreement number HR0011-20-9-0016 (PaPPa).

# A  Additional results

For completeness, additional plots obtained with different parameters are included. They are obtained using the following Quadrature strategies:

| Quadrature Alg | reltol | abstol | maxiters |
|---|---|---|---|
| CubaCuhre | $1 \times 10^{-4}$ | $1 \times 10^{-4}$ | 50 |
| HCubatureJL | $1 \times 10^{-4}$ | $1 \times 10^{-4}$ | 50 |
| CubatureJLh | $1 \times 10^{-4}$ | $1 \times 10^{-4}$ | 50 |
| CubatureJLp | $1 \times 10^{-4}$ | $1 \times 10^{-4}$ | 50 |

while the other strategies are the same as above (Section 4).

## A.1 Diffusion

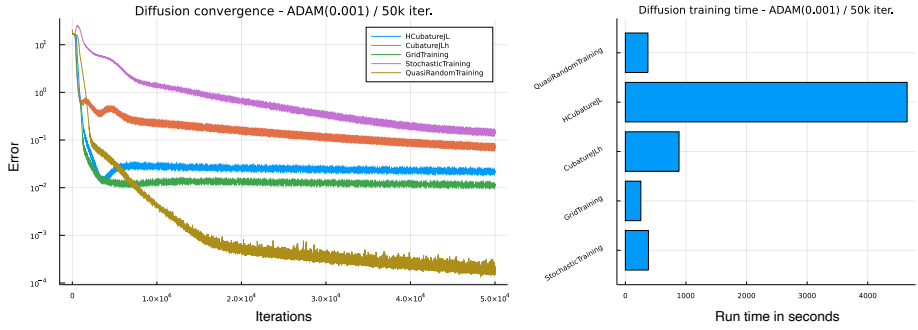

Figure 46: Convergence of the diffusion equation training using ADAM as minimizer.

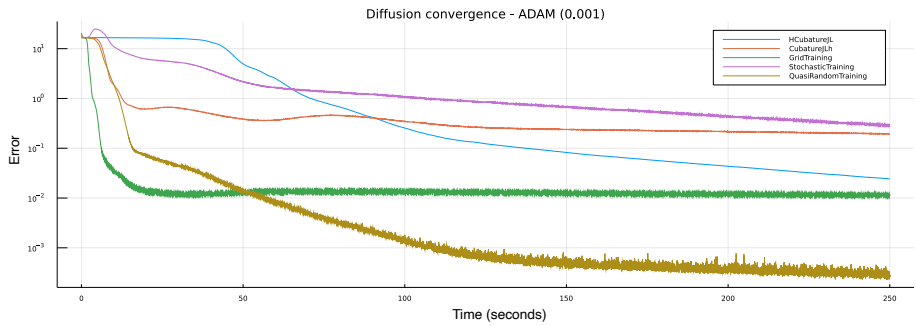

Figure 47: Training convergence from 0 to 250 seconds.

| Strategy | Minimum Error Value |
|---|---|
| HCubatureJL | $1.41 \times 10^{-2}$ |
| CubatureJLh | $7.02 \times 10^{-2}$ |
| GridTraining | $1.08 \times 10^{-2}$ |
| StochasticTraining | $1.42 \times 10^{-1}$ |
| QuasiRandomTraining | $1.32 \times 10^{-4}$ |

## A.2 Level Set

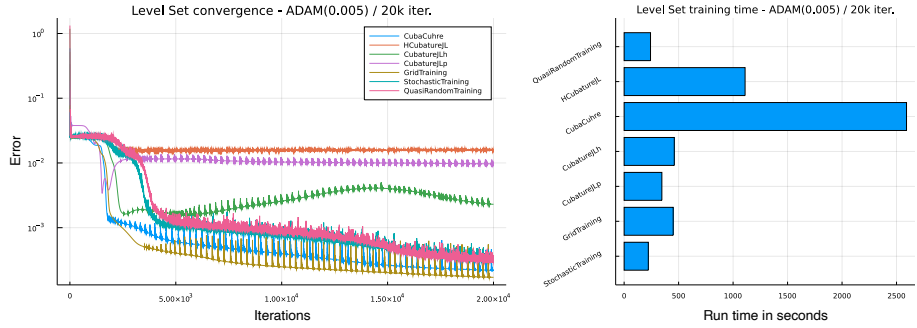

Figure 48: Convergence of the Level Set equation training using ADAM as minimizer.

| Strategy | Minimum Error Value |
|---|---|
| CubaCuhre | $2.19 \times 10^{-4}$ |
| HCubatureJL | $1.49 \times 10^{-2}$ |
| CubatureJLh | $1.52 \times 10^{-3}$ |
| CubatureJLp | $3.39 \times 10^{-3}$ |
| GridTraining | $1.71 \times 10^{-4}$ |
| StochasticTraining | $1.09 \times 10^{-4}$ |
| QuasiRandomTraining | $9.77 \times 10^{-5}$ |

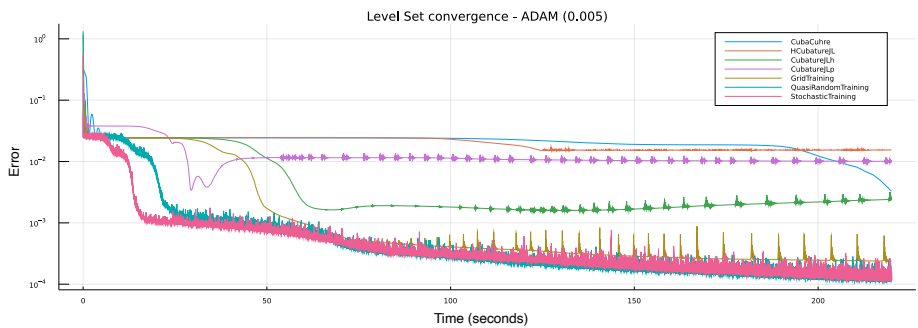

Figure 49: Training convergence with time on the horizontal axis (first 220 seconds of training).

## A.3 Allen Cahn

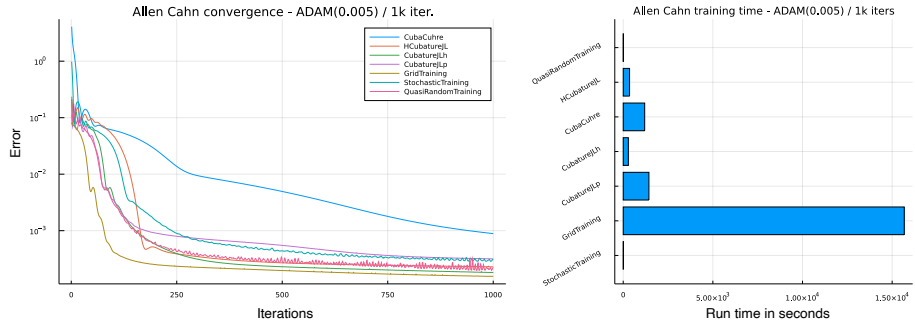

Figure 50: Convergence of the Allen Cahn equation training using ADAM as minimizer.

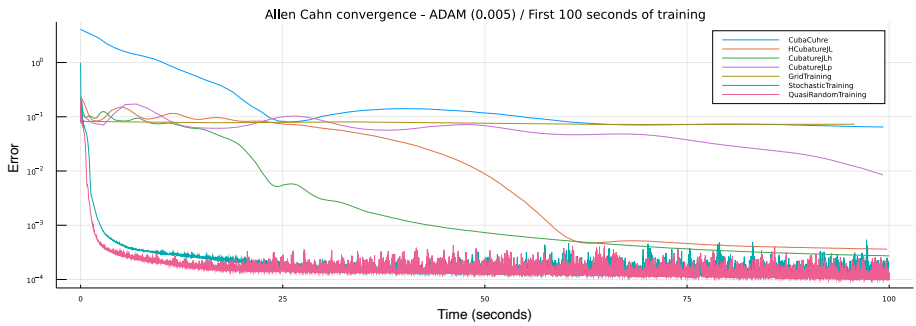

Figure 51: Training convergence with time on the horizontal axis (first 100 seconds of training).

| Strategy | Minimum Error Value |
| --- | --- |
| CubaCuhre | $8.87 \times 10^{-4}$ |
| HCubatureJL | $2.29 \times 10^{-4}$ |
| CubatureJLh | $1.82 \times 10^{-4}$ |
| CubatureJLp | $3.17 \times 10^{-4}$ |
| GridTraining | $1.55 \times 10^{-4}$ |
| StochasticTraining | $2.87 \times 10^{-4}$ |
| QuasiRandomTraining | $1.99 \times 10^{-4}$ |

## A.4 Hamilton Jacobi

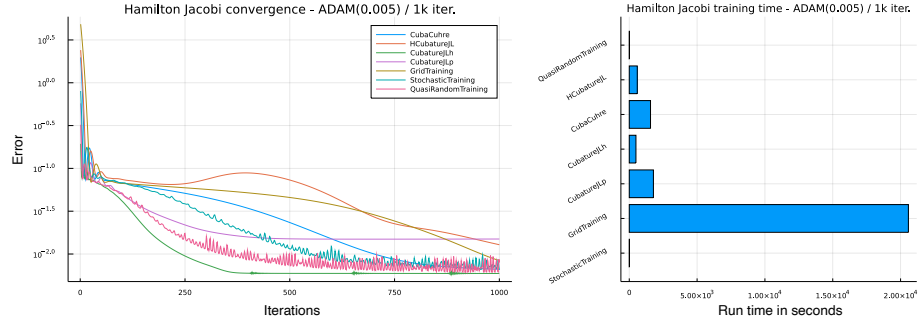

Figure 52: Convergence of the Hamilton Jacobi equation training using ADAM as minimizer.

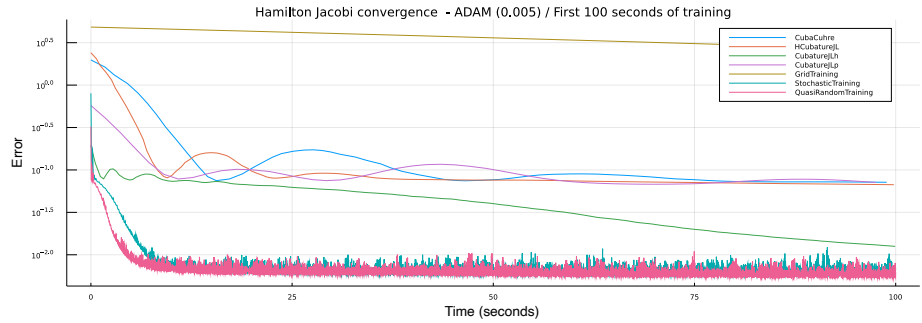

Figure 53: Training convergence with time on the horizontal axis (first 100 seconds of training).

| Strategy | Minimum Error Value |
| --- | --- |
| CubaCuhre | $6.64 \times 10^{-3}$ |
| HCubatureJL | $1.28 \times 10^{-2}$ |
| CubatureJLh | $5.83 \times 10^{-3}$ |
| CubatureJLp | $1.49 \times 10^{-2}$ |
| GridTraining | $8.38 \times 10^{-3}$ |
| StochasticTraining | $6.53 \times 10^{-3}$ |
| QuasiRandomTraining | $6.11 \times 10^{-3}$ |

[30] Wei-Liem Loh et al. On latin hypercube sampling. *Annals of statistics*, 24(5):2058–2080, 1996.

[31] Lu Lu, Xuhui Meng, Zhiping Mao, and George Em Karniadakis. Deepxde: A deep learning library for solving differential equations. *SIAM Review*, 63(1):208–228, 2021.

[32] E. Luján, H. Schinca, N. Olaiz, S. Urquiza, F.V. Molina, P. Turjanski, and G. Marshall. Optimal dose-response relationship in electrolytic ablation of tumors with a one-probe-two-electrode device. *Electrochimica Acta*, 186:494–503, December 2015.

[33] Yingbo Ma, Shashi Gowda, Ranjan Anantharaman, Chris Laughman, Viral Shah, and Chris Rackauckas. Modelingtoolkit: A composable graph transformation system for equation-based modeling. *arXiv preprint arXiv:2103.05244*, 2021.

[34] Scott G Marquis, Valentin Sulzer, Robert Timms, Colin P Please, and S Jon Chapman. An asymptotic derivation of a single particle model with electrolyte. *Journal of The Electrochemical Society*, 166(15):A3693–A3706, 2019.

[35] Wade S Martinson and Paul I Barton. A differentiation index for partial differential-algebraic equations. *SIAM Journal on Scientific Computing*, 21(6):2295–2315, 2000.

[36] Levi McClenny and Ulisses Braga-Neto. Self-adaptive physics-informed neural networks using a soft attention mechanism. *arXiv preprint arXiv:2009.04544*, 2020.

[37] William J Morokoff and Russel E Caflisch. Quasi-monte carlo integration. *Journal of computational physics*, 122(2):218–230, 1995.

[38] Balasubramanian Narasimhan, Manuel Koller, Steven G Johnson, Thomas Hahn, Annie Bouvier, Kiên Kiêu, Simen Gaure, and Maintainer Balasubramanian Narasimhan. Package 'cubature'. 2021.

[39] Sotirios E Notaris. An overview of results on the existence or nonexistence and the error term of gauss-kronrod quadrature formulae. *Approximation and Computation: A Festschrift in Honor of Walter Gautschi*, pages 485–496, 1994.

[40] H O'hara and Francis J Smith. Error estimation in the clenshaw–curtis quadrature formula. *The Computer Journal*, 11(2):213–219, 1968.

[41] Art B Owen. Quasi-monte carlo sampling. *Monte Carlo Ray Tracing: Siggraph*, 1:69–88, 2003.

[42] Guofei Pang, Lu Lu, and George Karniadakis. fpinns: Fractional physics-informed neural networks. *SIAM Journal on Scientific Computing*, 41:A2603–A2626, 01 2019.
75